\documentstyle[floats,eqsecnum,preprint,aps,psfig]{revtex}
\tightenlines

\def\Sv{\vec{S}}
\def\Sz{S^z}
\def\Jp{{J'}}
\def\deltap{\delta'}
\def\Jpc{{J'_c}}
\def\Jpeff{J'_{\rm eff.}}
\def\abs#1{\vert #1 \vert}
\def\Gap{{\sf E}}
\def\beq{\begin{equation}}
\newcommand{\eeq}[1]{\label{#1} \end{equation}}
\def\bea{\begin{eqnarray}}
\def\eea{\end{eqnarray}}
\def\nn{\nonumber}
\def\ie{{\it i.e.}}
\def\tr{{\rm tr}}

\font\amsmath=msbm10 scaled \magstep1
\font\amsmaths=msbm8
\def\Real{\hbox{\amsmath R}}
\def\real{\hbox{\amsmaths R}}

\def\Zed{\hbox{\amsmath Z}}

\let\rdraft=\draft

\catcode`\@=11
\def\marginnote#1{}
\newcount\hour
\newcount\minute
\newtoks\amorpm
\hour=\time\divide\hour by60
\minute=\time{\multiply\hour by60 \global\advance\minute by-\hour}
\edef\standardtime{{\ifnum\hour<12 \global\amorpm={am}%
        \else\global\amorpm={pm}\advance\hour by-12 \fi
        \ifnum\hour=0 \hour=12 \fi
        \number\hour:\ifnum\minute<10 0\fi\number\minute\the\amorpm}}
\edef\militarytime{\number\hour:\ifnum\minute<10 0\fi\number\minute}
\def\draftlabel#1{{\@bsphack\if@filesw {\let\thepage\relax
   \xdef\@gtempa{\write\@auxout{\string
      \newlabel{#1}{{\@currentlabel}{\thepage}}}}}\@gtempa
   \if@nobreak \ifvmode\nobreak\fi\fi\fi\@esphack}
        \gdef\@eqnlabel{#1}}
\def\@eqnlabel{}
\def\@vacuum{}
\def\draftmarginnote#1{\marginpar{\raggedright\scriptsize\tt#1}}
\def\draft{\oddsidemargin -.5truein
        \def\@oddfoot{\sl preliminary draft \hfil
        \rm\thepage\hfil\sl\today\quad\militarytime}
        \let\@evenfoot\@oddfoot \overfullrule 3pt
        \let\label=\draftlabel
        \let\marginnote=\draftmarginnote
   \def\@eqnnum{(\theequation)\rlap{\kern\marginparsep\tt\@eqnlabel}%
\global\let\@eqnlabel\@vacuum}  }
\def\underline#1{\relax\ifmmode\@@underline#1\else
        $\@@underline{\hbox{#1}}$\relax\fi}
\catcode`@=12
\relax

\rdraft

\begin{document}
\tolerance 50000
\author{D.C.\ Cabra$^{1}$\footnote{
On leave of absence from the Universidad Nacional de La Plata
and Universidad Nacional de Lomas de Zamora, Argentina.
}, A.\ Honecker$^{2}$\footnote{
A Feodor-Lynen fellow of the Alexander von Humboldt-foundation. \\
Present address: Institut f\"ur Theoretische Physik, TU Braunschweig,
     38106 Braunschweig, Germany.
}, P.\ Pujol$^{3}$
}
\address{
$^{1}$Physikalisches Institut der Universit\"at Bonn,
     Nu{\ss}allee 12, 53115 Bonn, Germany.\\
$^{2}$International School for Advanced Studies,
     Via Beirut 2-4, 34014 Trieste, Italy\\
{\em and} Institut f\"ur Theoretische Physik, ETH-H\"onggerberg,
     8093 Z\"urich, Switzerland. \\
$^{3}$Laboratoire de Physique\footnote{URA 1325 du CNRS associ\'ee
\`a l'Ecole Normale Sup\'erieure de Lyon.}
      Groupe de Physique Th\'eorique\\
      ENS Lyon, 46 All\'ee d'Italie, 69364 Lyon C\'edex 07, France.\\
}

\preprint{
\begin{minipage}[t]{\columnwidth}
\leftline{Eur.\ Phys.\ J.\ {\bf B13}, 55-73 (2000) \hfill cond-mat/9902112}
\rightline{BONN-TH-99-01, SISSA 10/99/EP,}
\rightline{ETH-TH/99-01, ENSL-Th 02/99}
\end{minipage}
}



\title{Magnetic Properties of Zig-Zag Ladders\footnote{
Work done under partial support of the EC TMR Programme
{\em Integrability, non-per\-turba\-tive effects and symmetry in
Quantum Field Theories}, grant FMRX-CT96-0012.
}}
\date{Februray 9, 1999; revised May 21, 1999; final modification June 16, 1999}
\maketitle
\begin{abstract}
\begin{center}
\vspace*{-9mm}
\parbox{15cm}{\advance\baselineskip-1pt
We analyze the phase diagram of a system of spin-$1/2$ Heisenberg
antiferromagnetic chains interacting through a zig-zag
coupling, also called zig-zag ladders. Using bosonization techniques
we study how a spin-gap or more generally plateaux in magnetization
curves arise in different situations. While for coupled $XXZ$
chains, one has to deal with a recently discovered chiral
perturbation, the coupling term which is present for normal ladders
is restored by an external magnetic field, dimerization or the presence
of charge carriers.
We then proceed with a numerical investigation of the phase diagram
of two coupled Heisenberg chains in the presence of a magnetic field.
Unusual behaviour is found for ferromagnetic coupled antiferromagnetic
chains. Finally, for three (and more) legs one can choose
different inequivalent types of coupling between the chains.
We find that the three-leg ladder can exhibit a spin-gap and/or
non-trivial plateaux in the magnetization curve whose
appearance strongly depends on the choice of coupling.
}
\end{center}
\end{abstract}
\vspace{0.2cm}
\pacs{
\hspace{-13mm}
}

\newpage

\section{Introduction}

In the last few years, the study of quasi-one dimensional
magnets has become intense. One of the main reasons is the
appearance of real materials which can be well approximated
by one-dimensional models. An important class corresponds
to the so-called spin ladder materials, such as the compounds
Sr$_{1-x}$Cu$_x$O$_2$ and La$_{4+4x}$Cu$_{8+2x}$O$_{14+8x}$
which are closely related to high-$T_c$ compounds
(for reviews see e.g.\ \cite{review,Ptoday}) or the organic
two-leg ladder material Cu$_2$(C$_{5}$H$_{12}$N$_{2}$)$_2$Cl$_4$
(see e.g.\ \cite{CJYFHBLHP}).

In addition, the similarities in the normal-state properties
of high-temperature superconductors and ladder cuprates
make these latter systems valuable laboratories from
both the theoretical and experimental point of view.

Due to the low dimensionality, quantum fluctuations are crucial
and because of this, these systems exhibit a variety of interesting
phenomena such as the appearance of plateaux in magnetization curves,
an issue that has received a lot of attention recently (see e.g.\
\cite{Hida,Tone,AOY,CCLMMP,HLP,Totsuka,CHP,SaTaS3o2,Totsuka2,CHP2,CG,%
H,Mila,TLPRS,Totsuka3,FZ,CG2}).

The study of spin ladder systems with different topologies of
couplings has been also motivated both from the experimental
and the theoretical side. In particular, the zig-zag coupling between
quantum spin chains has received much theoretical attention
\cite{CPKSR,WA,AS,NGE,Sor}. In the case of
antiferromagnetic (AF) couplings, the zig-zag array introduces frustration
which makes the study of these systems much more difficult.
Apart from being a possible approach to study the two-dimensional triangular
lattice, this topology of couplings is realized in a number
of quasi-one dimensional compounds, such as Cs$_2$CuCl$_4$ \cite{CTC},
KCuCl$_3$ and TlCuCl$_3$ \cite{STKTKTMG} as well as NH$_4$CuCl$_3$ \cite{STKT}.
A (two-dimensional) zig-zag arrangement is also present in SrCuO$_2$.
At room temperature, the materials KCuCl$_3$, TlCuCl$_3$ and
NH$_4$CuCl$_3$ are isostructural and can be described by an
alternating two-leg zig-zag ladder \cite{STKTKTMG,STKT}. Experimentally,
one observes a zero magnetization plateau in the low-temperature
magnetization process of KCuCl$_3$ and TlCuCl$_3$ \cite{STKTKTMG}, whereas
for NH$_4$CuCl$_3$ plateaux in the magnetization curve are observed
at $1/4$ and $3/4$ of the saturation magnetization \cite{STKT}. While the
former is in good agreement with the theoretical predictions (which will
be discussed in detail in this paper), the explanation of the latter
is still unclear (even though explanations have already been
proposed \cite{Kolezhuk}). A detailed analysis of related structures
such as those studied in the present paper can be expected to
be also useful for interpreting these experiments.

Zig-zag coupled chains also have several special properties which render
them an interesting problem from a purely theoretical point of view.
Firstly, unlike many other systems they do not have a simple
`strong-coupling' limit where the system is decoupled into finite clusters
of spins. If there is such a decoupling limit, the possible values of the
magnetization are clearly quantized in this limit and using perturbation
theory one can easily understand the appearance of
magnetization plateaux in the strong-coupling region \cite{CHP,CHP2,H}.
For example, for the usual spin ladders, one finds a quantization
condition on the magnetization $\langle M \rangle$ (which we normalize to
saturation value 1) for the appearance of a plateau in the magnetization
curve \cite{CHP,CHP2} (compare also \cite{AOY,Totsuka,CG,H,Totsuka3}):
\beq
S V (1 - \langle M \rangle) \in \Zed \, .
\eeq{condM}
Here $S$ is the spin on each site and $V$ is the volume of a
translationally invariant unit cell. For $N$-leg ladders,
$V = l N$ where $l$ is the period of explicit or spontaneous
breaking of translational symmetry in the magnetized groundstate.

Since zig-zag coupled chains with no or weak dimerization do not
have a simple strong-coupling limit,
it is not immediately clear if the condition for the appearance of
plateaux in zig-zag ladders will also be given by (\ref{condM}).
The fact that the zig-zag coupling is frustrating is a further
reason why the quantization condition on the magnetization or e.g.\ the
universality classes of the transitions at the plateau boundaries
might be different. From a field theoretical point of view, zig-zag
coupling is also interesting because at zero field it cancels the
most relevant coupling term for the usual ladders and instead
one has to deal with a chirally asymmetric perturbation \cite{NGE}.

After mentioning all these possibilities for a different behaviour
we should, however, immediately point out that one of
our main conclusions is going to be that zig-zag coupling is
not very different from the ordinary one. In particular, the
most relevant interaction term is recovered in many situations like
the presence of a magnetic field, dimerization or charge carriers.
We will also find that all observed magnetization plateaux have a
natural interpretation in terms of the quantization condition (\ref{condM}).

Motivation for studying these systems arises also from other fields.
For example, spin ladders arise in the study of gated Josephson
junction arrays \cite{AlAu}. Even more, Ref.\ \cite{WA} pointed
out a close analogy between the two-leg zig-zag ladder and the
doped Kondo lattice model. This model in turn is believed to
be relevant to the phenomenon of colossal magnetoresistance in
metallic oxides and has therefore received renewed attention
recently \cite{YHMM}.

\begin{figure}[hpt]
\psfig{figure=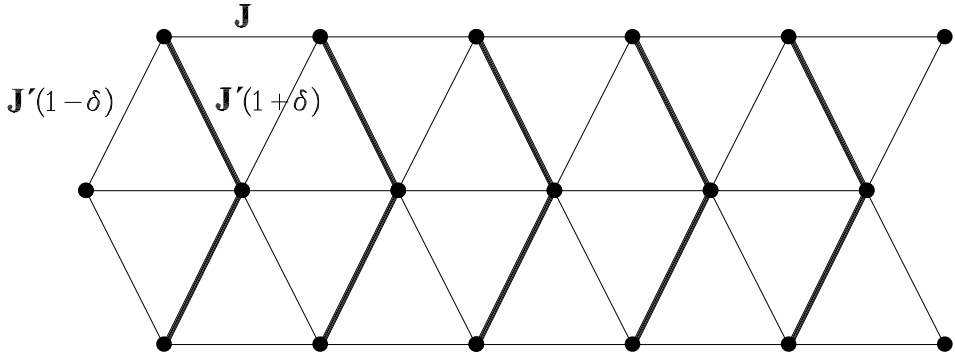,width=\columnwidth,angle=0}
\smallskip
\caption{
Generic structure of the ladders considered in the present paper.
\label{figdim1}
}
\end{figure}

The main focus of the present paper are $N$ coupled Heisenberg
chains with a dimerized zig-zag coupling (see also Fig.~\ref{figdim1})
\beq
H^{(N)} = J \sum_{i=1}^N \sum_{x=1}^L \Sv_{i,x} \cdot \Sv_{i,x+1}
+ \Jp \sum_{i=1}^N \sum_{x=1}^L \Sv_{i,x} \cdot \left(
\left(1 + \delta\right) \Sv_{i+1,x} +
\left(1 - \delta\right) \Sv_{i+1,x+1} \right)
- h \sum_{i,x} \Sz_{i,x} \, .
\eeq{zzHam}
For compactness of presentation we have written scalar products here,
but we will also consider the case where an $XXZ$ anisotropy $\Delta$
is introduced in the obvious way. We will furthermore discuss
a similar system including dimerization along the chains
and also charge degrees of freedom, \ie\ Hubbard zig-zag ladders.
In eq.\ (\ref{zzHam}) there is also some ambiguity in writing the
interchain coupling, in particular in combination with specifying
boundary conditions for $N>2$ -- an issue to which we shall return
later.

The two-leg zig-zag ladder can be reinterpreted as a Heisenberg
chain with next-nearest-neighbour interactions. In this interpretation,
the study of this system has a long history going back at least to
\cite{MaGo}. One intriguing observation regarding this system is that
for special parameters the groundstate takes a very simple form
and correlation functions can be computed exactly \cite{Maju,ShSu}.

This paper is organized as follows: In Section II we
examine the Hamiltonian (\ref{zzHam}) at $\delta=h=0$ without and
with $XXZ$ anisotropy, using non-Abelian and Abelian bosonization,
respectively. Section III is devoted to a field-theoretical
analysis of the effect of various modifications: An external
magnetic field, dimerization or doping with charge carriers.
In Section IV we then numerically investigate the magnetization
process of the two-leg ladder (\ref{zzHam}) and discuss the
various commensurate and incommensurate phases appearing for
$\Jp > 0$ as well as $\Jp < 0$. In Section V we investigate the
transition to saturation of the two-leg zig-zag ladder in some
detail. Finally, in Section VI we numerically compute magnetization
curves for several variants of the three-leg zig-zag ladder.
An appendix contains some supplementary material.


\section{Field Theory for Zero Magnetic Field}

\subsection{The $SU(2)$ symmetric case}

First, we reexamine the simplest case of the $SU(2)$ symmetric
two-leg zig-zag ladder that has been studied previously in \cite{WA,AS}.

We start by reviewing some of the earlier results.
In  the weak (interchain) coupling limit each of the
chains can be described by the level 1 Wess-Zumino-Witten (WZW) model
\cite{Wi,A2} with the action given by
\beq
W[g^i]=
\frac{1}{8\pi} \int_{\real^2} d^2x
\tr(\partial_{\mu}g^i\partial^{\mu}{g^i}^{-1})
+ \Gamma[g^i],
\eeq{1'}
where the superscript $i$ labels the different chains,
$g^i$ takes values in the Lie group $SU(2)$ and $\Gamma[g]$ is
the Wess-Zumino term
\beq
\Gamma[g]=\frac{1}{12\pi} \int_{Y} d^3y \epsilon_{\alpha \beta \gamma}
\tr(g^{-1}\partial_\alpha g g^{-1}\partial_\beta g g^{-1}\partial_\gamma g),
\eeq{wzt}
with $Y$ a three-dimensional manifold having $\Real^2$ as boundary.

The bosonized expression for the spin operator is given by
\beq
\vec S^i(x)=\vec J^i_{R}+\vec J^i_{L}+B(-1)^x \tr(\vec \sigma g^i)
\eeq{spin}
where $J^{i,a}_{R} = -\frac{1}{2\pi} \tr((\partial_+g^i)(g^i)
^{-1}\sigma^a)$,
$J^{i,a}_{L} = \frac{1}{2\pi} \tr((g^i)^{-1}\partial_-g^i \sigma^a)$ are the
WZW currents satisfying a $\widehat{su(2)}_1$
Kac-Moody algebra and $B$ is a non-universal constant.

The marginally irrelevant perturbation term
\beq
- \lambda \sum_i \int dx \vec J^i_R \cdot \vec J^i_L~,~~~~~~\lambda > 0,
\eeq{marPer}
which is responsible for the logarithmic corrections
in the correlators in the case of a single spin chain
is usually discarded when there are more relevant terms
(such as those arising in the normal ladders from the interchain
coupling). In the present case, though, it should be kept since the
zig-zag interchain coupling gives rise only
to marginal perturbations.

More precisely, the zig-zag coupling
\beq
H_{{\rm int}} = \Jp \sum_{x} \vec S^1_x \cdot (\vec S^2_x + \vec S^2_{x+1}),
\eeq{zz}
leads to the (marginal) current-current interaction
\beq
H_{{\rm int}} = \alpha \int dx (\vec J_R^1+\vec J_L^1)
        \cdot (\vec J_R^2+\vec J_L^2),
\eeq{int}
where $\alpha$ is small and proportional to the interchain coupling $\Jp$.
It is positive for antiferromagnetic coupling and negative for a
ferromagnetic coupling.
Discarding non-Lorentz invariant terms, which do not contribute to the
one-loop renormalization group (RG) equations, one can rewrite the Hamiltonian
in the following way
\beq
H = H_0^1 + H_0^2 - \beta \sum_{i=1}^2 \int dx (\vec J_R^i \cdot \vec J_L^i)
+ \alpha \int dx \vec J_R^T \cdot \vec J_L^T
\eeq{totham}
where $\beta = \lambda + \alpha$ and $\vec J^T = \vec J^1 + \vec J^2$
which satisfies a level 2 Kac-Moody algebra.
Equation (\ref{totham}) is the Hamiltonian considered in \cite{WA,AS}.
By a one-loop RG analysis one finds an exponentially small gap
for antiferromagnetic zig-zag coupling and a massless regime for
a ferromagnetic coupling. However,
in \cite{NGE} it was shown that there is another (marginal) term
\beq
\gamma\int dx \left( \tr( \vec \sigma g^1)\cdot  \partial_x \tr(  \vec
\sigma g^2) -
\tr( \vec \sigma g^2)\cdot  \partial_x \tr(  \vec \sigma g^1) \right),
\eeq{pbsu2}
with $\gamma \propto \Jp$, that could change the behaviour of
the system in the ferromagnetic regime.

Now we proceed to discuss how the RG equations of \cite{AS} are
changed when the chirally asymmetric perturbation term (\ref{pbsu2})
is included in the non-Abelian bosonization framework. Such a
computation was actually already performed in \cite{NGE} using a
fermionic representation.  Still, our result will be complementary
to \cite{NGE} and is specially appropriate to analyze the $SU(2)$ symmetric
case and generalize it to $N > 2$ since this symmetry is explicit in
the present analysis.

It is easy to see that another $SU(2)$ invariant operator given by
\beq
\gamma' \int dx \left( \tr( g^1)  \partial_x \tr( g^2) -
\tr( g^2)  \partial_x \tr( g^1) \right),
\eeq{pb2u2}
has to be included as a counterterm in the action. Using the operator
product expansion of these operators and a rescaling of the coupling
constants, one obtains the RG equations:
\bea
{d\alpha \over d\ln \ell} &=&  \alpha^2 + \gamma^2
\, ,  \nn \\
{d\beta \over d\ln \ell} &=& - \beta^2 + 2 \alpha \beta
+ \gamma^2 \, , \nn \\
{d\gamma \over d\ln \ell} &=& 2 \gamma \alpha + 2 \gamma \beta
                  + 2 \gamma' \alpha \, , \nn \\
{d\gamma' \over d\ln \ell} &=& 6 \gamma \alpha
                           - 6 \gamma' \beta + 6 \gamma' \alpha
\, .
\label{RGeq2}
\eea
The RG equations obtained in \cite{AS} can be recovered from here
by just putting $\gamma = \gamma' = 0$. It can be shown that
for an AF coupling the presence of these $\gamma$ and $\gamma'$
terms does not affect the qualitative behaviour of the flow.
The system is driven to a strong coupling regime, corresponding to
the massive behaviour mentioned in \cite{AS}. The way terms like
(\ref{pbsu2}) affect the nature of the massive excitations in
this strong coupling regime is nevertheless not so clear.

These equations can be generalized to an arbitrary number of coupled
chains, provided all the chains are coupled in the same way.
This applies in particular to $N=3$ with periodic boundary conditions
(PBC). One can then conclude that
three chains with a periodic arrangement of antiferromagnetic
coupling $\Jp > 0$ are massive at least in the weak-coupling region
$\Jp \ll J$. Due to the dimensions of the perturbing operators involved,
this kind of coupling can be considered as weaker than
the normal coupling, in the sense that the growth of the associated
coupling is governed by a smaller exponent in the present case.
This shows the importance of a careful
treatment of the marginal interactions. This should help to clarify
the problem of the presence of a gap in the weak-coupling region
for the usual $N=3$ ladder with PBC.

The situation for a ferromagnetic coupling is more subtle. One can see that
the trivial massless infrared fixed point $\alpha = \beta =0$ is stable only if
$\gamma = \gamma' =0$. Then the presence of the twist term affects the large
scale behaviour of the system, preventing it to reach the trivial massless
point. So, the one-loop RG flows for weak $\Jp < 0$ and $\Jp > 0$ become
similar. This has already been noticed in \cite{NGE} where it was
argued that the dimerized phase seems to extend into the
ferromagnetic region. One should however keep in mind that this is just
a one-loop calculation and non-perturbative effects could change the
large scale behaviour of the system.

We conclude the present section stressing that all these results are
easily generalizable to the weak-coupling regime of an arbitrary number
$N$ of coupled chains, provided that in the continuum limit
all chains are coupled in an equal manner. It should be noted,
however, that on the lattice, inequivalent versions of completely symmetric
zig-zag interchain coupling exist for $N > 2$. After taking the continuum
limit, these differences manifest themselves in different signs for
interaction terms of the type (\ref{pbsu2}), \ie\ most of the originally
completely symmetric boundary conditions are not symmetric anymore after taking
the continuum limit. Note also that the cases of PBC where not all
pair-couplings are present (which is the case for $N>3$) or open
boundary conditions (OBC) for $N>2$
are much more subtle and it is not clear if the above results apply also
to them.

\subsection{The $XXZ$ case}

Let us consider now the addition of an $XXZ$ anisotropy.
In this case, the $SU(2)$ symmetry of the system
is broken and it is then more appropriate to adopt the
Abelian bosonization approach. Below we follow the notations
of \cite{A2,WA} in order to simplify making the connection with
the non-Abelian description in terms of WZW fields of the previous
section (this implies some minor changes of conventions such as
a rescaling of the fields with respect to our study of the
ordinary spin ladders \cite{CHP2}).

In the Abelian bosonization language, each chain is described
by a compactified free bosonic field $\phi^i$ with its dynamics governed by
\footnote{Note that a factor $(4R)^{-2}$ is missing in front of
the $\Pi^2$ term in eqs.\ (2.2) and (3.4) of \cite{CHP2}.}
\beq
H = {1 \over 2} \int dx \left( v K (\partial_x \tilde{\phi})^2
+ {v \over K} (\partial_x \phi)^2 \right)
\eeq{Ham}
The field $\phi^i$ and its dual $\tilde{\phi}^i$ are given by the
sum and difference of the lightcone components, respectively. The
constant $K$ governs the conformal dimensions of the bosonic vertex
operators and can be obtained exactly from the Bethe ansatz
solution of the $XXZ$ chain (see e.g.\ \cite{CHP2} for a detailed
summary). We have $K=1$ for the $SU(2)$ symmetric case ($\Delta =
1$) and it is related to the radius $R$ of \cite{CHP2} by $K^{-1} =
2 \pi R^2$.

In terms of these fields, the spin operators read
\bea
S_{i,x}^z &=& {1 \over \sqrt{2\pi}} \partial_x \phi^i
+ a : \cos(2 k_F^i x + \sqrt{2 \pi} \phi^i): + \frac{\langle M^i
\rangle}{2} \, , \label{sz} \\
S_{i,x}^{\pm} &=& (-1)^x
:{\rm e}^{\pm i\sqrt{2\pi} \tilde{\phi}^i}
\left(b \cos(2 k_F^i x + \sqrt{2 \pi} \phi^i) + c \right) : \, ,
\label{s+}
\eea
where the colons denote normal ordering with respect to the
groundstate with  magnetization $\langle M^i\rangle$. The Fermi
momentum $k_F^i$ is related to the magnetization of the $i$th chain
as $k_F^i = (1-\langle M^i \rangle )\pi/2$. The effect of an $XXZ$
anisotropy and/or the external magnetic field (to be discussed
later) is then to modify the scaling dimensions of the physical
fields through $K$ and the commensurability properties of the spin
operators, as can be seen from (\ref{sz}), (\ref{s+}). The
non-universal constants $a$, $b$ and $c$ can be in general computed
numerically (see e.g.\ \cite{HF}, for the case of zero magnetic
field) and in particular the constant $b$ has been obtained exactly
in \cite{LZ}.

At zero magnetization, (\ie\ $k_F^i = \pi /2$), the first and second terms
in each of these equations correspond to the components of the non-oscillatory
($\vec J^i_R+\vec J^i_L$)  and oscillatory ($\tr(\vec \sigma g^i)$)
terms in (\ref{spin}).

Let us consider now a two-leg zig-zag $XXZ$ ladder.
The perturbation terms in the case
of zero magnetization can be separated into three classes:

\begin{itemize}
\item[i)]
Terms quadratic in the derivatives
\beq
\alpha \partial_x \phi^1(x)   \partial_x \phi^2(x) ,
\eeq{radiusren}
that can be absorbed into a renormalization of the compactification radii,
once we diagonalize the derivative part going to the new variables
$\phi^{\pm}=(\phi^1\pm \phi^2)/\sqrt{2}$.
The $K$ parameters are then renormalized as
\beq
K_{\pm} = K (1 \mp 2 \Jp K/(J \pi) + O((\Jp/J)^2)) .
\eeq{rad}
\item[ii)]
The other contributions from current-current intrachain and interchain
interactions can be rewritten as
\bea
&-\lambda & ~~\cos (\sqrt{4\pi} \phi^+) \cos (\sqrt{4\pi} \phi^-)
\nonumber\\
&+\alpha &~~\cos (\sqrt{4 \pi} {\tilde{\phi}}^-) \left( \cos (\sqrt{4\pi} \phi^+)+
\cos (\sqrt{4\pi} \phi^-)\right).
\label{pop}
\eea
The dimensions of these operators are given by
\beq
K_+ + K_-  \ \ \ \hbox{and} \ \ \
{1 \over K_-} + K_+,
\eeq{dimOp}
respectively (the third operator in (\ref{pop}) is always irrelevant).
The $\lambda$ term (which is the standard current-current term for
the individual chains) is irrelevant for $\Delta < 1$, while
the $\alpha$ term is
irrelevant for a ferromagnetic interchain coupling, except for $K =1$
(and $\Jp = 0$)
which corresponds to the $SU(2)$ symmetric point where it is marginal.
For an AF coupling, the $\alpha$ term is relevant for $\Delta \ge 1$.
\item[iii)]
We also have chirally asymmetric terms \cite{NGE}
(those corresponding to (\ref{pbsu2})) which now read
\beq
-\partial_x \phi^+(x) \sin (\sqrt{4\pi} \phi^-)
\, ; \quad + \partial_x \phi^-(x) \sin (\sqrt{4\pi} \phi^+)
\, ; \quad - \partial_x {\tilde \phi}^+(x) \sin (\sqrt{4\pi} {\tilde\phi}^-)
 \, .
\eeq{pbt}
Their dimensions are
\beq
1+ K_- ,\ \ \
1+ K_+ ,\ \ \ \hbox{and}
\ \ \ 1+ {1 \over K_-} \, .
\eeq{dimOp2}
If we consider a small $XXZ$ anisotropy with $\Delta \ll 1$, both $K_{\pm}$
increase and the term
$$ \partial_x {\tilde \phi}^+(x) \sin (\sqrt{4\pi} {\tilde\phi}^-)$$
will now become the most relevant for sufficiently small $\Jp$.

\end{itemize}

Following arguments similar to the ones of \cite{NGE} it follows that
at most the symmetric ($\phi^+$) sector is massless. For
$\Delta > 1$, also this field acquires a mass, but stays
massless for $\Delta \ll 1$. The above analysis cannot directly be
applied to the region $\Delta \approx 1$, because higher loops should
be included. Moreover, at the point $\Delta =1$, the $SU(2)$
symmetry is not directly explicit in this treatment. However,
according to the one-loop RG analysis in \cite{NGE} and Section II.A of
the present paper, a gap seems to open at $\Delta = 1$ for small
$\abs{\Jp} \ne 0$, regardless of its sign.

\section{Stability of the Zig-Zag Coupling and Magnetization Plateaux}

In the present section we will study different  mechanisms
due to which the zig-zag interchain coupling becomes unstable against
the perpendicular interchain coupling.

The outcome is that only under very special circumstances
the zig-zag coupling is stable, while  under the action of an external
magnetic field, the addition of rung dimerization or the presence of
charge carriers, an effective perpendicular coupling is generated.

\subsection{$XXZ$ in a magnetic field}

When we add an external magnetic field to the $XXZ$ zig-zag ladder,
the situation is more subtle, since the staggered terms do not
necessarily cancel. Indeed, we show in this section that new non-oscillating
commensurate terms arise.

The current-current terms become
\beq
\sin k_F \cos(\sqrt{4 \pi} {\tilde{\phi}}^-)\left( \sin k_F \cos
(\sqrt{4\pi} \phi^++4k_F x)+
 \sin(k_F-\sqrt{4\pi} \phi^-)\right) ,
\eeq{ccm}
and the other terms become
\bea
-\partial_x \phi^+(x) \sin (2k_F-\sqrt{4\pi} \phi^-) - \frac{1}{\sqrt 2}
\left(
\partial_x \phi^2(x) \sin (4k_F x + 2k_F + \sqrt{4\pi} \phi^+)\right.
\nonumber \\
\left. -\partial_x \phi^1(x) \sin (4k_F x - 2k_F + \sqrt{4\pi} \phi^+)\right)
-\partial_x {\tilde \phi}^+(x) \sin (\sqrt{4\pi} {\tilde\phi}^-) \, .
\label{pbm}
\eea

The main difference with the case at $\langle M \rangle =0$ is that now
we have a very relevant term which is proportional to
\beq
\cos ^2 k_F \cos (4k_F x + \sqrt{4\pi} \phi^+) + \cos  k_F \cos (k_F  -
\sqrt{4\pi} \phi^-) \eeq{rt}
with dimensions $K_+$ and $K_-$.
Note that the first term in (\ref{rt}) will disappear since it is
incommensurate for $\langle M \rangle \ne 0$.
Hence this term gives a mass to the $\phi^- $ field, leaving
only the symmetric field massless.

We can then make the following statement about the phase diagram of
two zig-zag coupled $XXZ$ spin chains with $\Delta < 1$:
For non-zero magnetization, and in all cases, one of the degrees of freedom
is massive and the other massless, leaving a $c=1$ theory.

With some modifications to the argumentation of \cite{CHP2},
this can be easily generalized to $N$ weakly coupled zig-zag chains,
provided all the chains are coupled together:
The generalization of the most relevant
interaction term (\ref{rt}) to the case of $N$-leg ladders is:
\beq
\sum_{i,j} \left\{ \lambda_1 \cos^2{k_F}
    : \cos(4 x k_F + \sqrt{2\pi}(\phi_i + \phi_{j})) :
+ \lambda_2 \cos{k_F} : \cos(k_F - \sqrt{2\pi} (\phi_i - \phi_{j})) :
 \right\} \, .
\eeq{pert2}
As for the case $N=2$, the coupling constants $\lambda_i$ essentially
correspond
to the coupling $\Jp$ between the chains: $\lambda_i \sim \Jp/J$, but have a
non-trivial dependence on $\langle M \rangle$:
$\lambda_i \rightarrow 0$ for $\langle M\rangle \rightarrow 0$.
The Gaussian part of the Hamiltonian is now given by:
\beq
\bar{H}^{(N)} = \int {\rm d}x \Biggl[
      {v \over 2} \sum_{i=1}^N \left\{
 K \left(  \partial_x \tilde{\phi}^i(x)    \right)^2 +
   {1\over K} \left(\partial_x \phi^i(x)\right)^2
\right\}
 + {\lambda \over  \pi} \sum_{i,j} \left(\partial_x \phi_i(x)\right)
      \left(\partial_x \phi_{j}(x)\right) \Biggr] \, ,
\eeq{LeH}
where $\lambda \sim \Jp/J$. As for the case of the normal coupled ladder,
the last term produces a shift of the compactification radii of the fields
and plays a crucial r\^ole in the opening of non-trivial plateaux.
In arriving to the Hamiltonian (\ref{LeH}) we have discarded a constant term
and absorbed a term linear in the derivatives of the free bosons into a
redefinition of the applied magnetic field. One has also to include the
generalization
to $N$ chains of the current-current term and the twist term.
While obtaining the expression for the first one is an easy task, writing down
the twist term for generic $N$ has some subtleties.
Keeping only commensurate terms, a na\"{\i}ve generalization of (\ref{pbt})
would be:
\bea
\sum_{i>j}
- \partial_x (\phi_i + \phi_j) \sin (2k_F - \sqrt{2\pi} (\phi_i - \phi_j))
- \partial_x (\tilde{\phi}_i + \tilde{\phi}_j)
\sin (\sqrt{2\pi} (\tilde{\phi}_i - \tilde{\phi}_j)) \, .
\label{genpbt}
\eea
If we take into account only the most relevant perturbation term
(\ref{pert2}),
the analysis is similar to the one of the normal ladder given in
\cite{CHP2}. In particular, one can radiatively generate ``$N$-Umklapp''
terms like
$\cos\left(2x\sum_{i=1}^{N} k_{F}^i + \sqrt {4\pi N} \psi_D \right)$,
where $\psi_D = {1\over \sqrt{N}} \sum_{i=1}^{N} \phi_i$,
which can give rise to the appearance of a plateau if
(\ref{condM}) is satisfied with $V=N$, $S=1/2$.

The zero-loop analysis for the opening of such plateaux is basically
the same as in \cite{CHP2}. Just the coefficient of the interaction
arising from the smooth part acquires an extra factor of two.
Via the Gaussian part of the Hamiltonian this replaces $\Jp$
by $2 \Jp$ in the dimension formula for the $N$-Umklapp term.
In this way, we obtain for example,
at $\Delta = 1$, $\Jpc \approx 0.045 J$ for the
$\langle M\rangle = 1/3$ plateau at $N=3$ and
$\Jpc \approx 0.35 J$ for $\langle M \rangle = 1/2$ at $N=4$
and also for $\langle M \rangle = 1/5$ at $N=5$.
At this level one could expect the magnetization curves of zig-zag
ladders to be very similar to the ones of normal ladders.
There are however some subtleties which are not captured by this analysis.
There are inequivalent ways of coupling
a number $N\ge 3$ chains in a periodic zig-zag way.
These inequivalent ways of
PBC couplings correspond in the weak coupling field theoretical
model to changing the relative signs, or what is equivalent, permuting
chain indices in the expression of the twist term (\ref{genpbt}).
Since the zero-loop treatment presented above does not take into account
the twist term, it is obvious that we have to consider loop corrections
to take into account this effect.
A detailed RG treatment of this
model taking into account the current-current and twist term is a difficult
task beyond the scope of this paper. However, exact diagonalization of
finite chains for a strong enough $\Jp / J$ ratio will show that
these inequivalent couplings can give rise to different behaviour in the
presence of a magnetic field.
Another problem is the extension of these results to OBC where we encounter
the same limitations as for the normal ladders. We refer the reader to
\cite{CHP2} for a detailed discussion of this point. We just mention here
that the results above cannot be directly extrapolated to configurations with
OBC. In this case one should therefore complement the present field
theoretical analysis by other methods such as exact diagonalization.

\subsection{Dimerization in $XXZ$ ladders}

Another way to generate an effective perpendicular coupling is to include
dimerization in the zig-zag coupling. Let us consider the Hamiltonian in
(\ref{zzHam}) with non-zero rung dimerization  $\delta$. In the bosonized
language, this gives rise to the perturbation term
\bea
H_{\rm dim} &=& \Jp \sum_i
  \sum_x \left((-1)^x \vec n_i \cdot \left( (1+\delta )
(-1)^x \vec n_{i+1} +
(1-\delta ) (-1)^{x+1} \vec n_{i+1} \right) \right) \nn \\
&=& 2 \Jp \delta \sum_i \sum_x \vec n_i \cdot \vec n_{i+1} \, ,
\label{dim}
\eea
where $\vec n_i$ is the staggered component of the spin operator
at zero magnetization ($n^z_i = \cos(\sqrt{2 \pi} \phi^i)$,
$n^{\pm}_i = \exp(\pm i \sqrt{2 \pi} \tilde{\phi}^i)$).

This perturbation term can be rewritten as
\bea
H_{\rm dim} = \lambda \sum_i \sum_x
&&\left[ - \cos \left(\sqrt{4\pi} (\phi_i+\phi_{i+1})\right)
   + \cos \left(\sqrt{4\pi}  (\phi_i-\phi_{i+1})\right) \right. \nn \\
&& \left.
+ 2 \cos \left(\sqrt{4\pi} (\tilde\phi_i-\tilde\phi_{i+1})\right) \right] \, ,
\label{Hdim}
\eea
where $\lambda \propto \Jp \delta$.

The effective model is then very similar to the one for a ladder with a
normal perpendicular coupling. In particular, using the formulae
of \cite{CHP2}, one can study the opening of plateaux in the magnetization
curve as a function of $\delta$ and $\Jp$. The $N=2$ version
of the dimerized zig-zag ladder has been analyzed in detail in \cite{Totsuka2}
and the magnetization plateau with $\langle M \rangle = 1/2$ predicted
there was also observed numerically \cite{TNK,FGKMW}.

\begin{figure}[hpt]
\psfig{figure=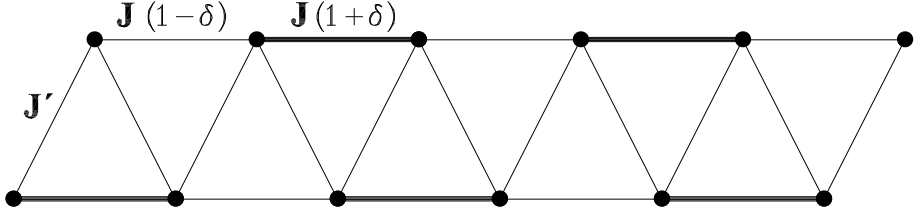,width=\columnwidth,angle=0}
\smallskip
\caption{
Dimerization along the legs.
\label{figdim2}
}
\end{figure}

Another way to add dimerization is along the legs of the chains
as in Fig.\ \ref{figdim2}, which gives rise to additional terms.
They read in the presence of a magnetic field
\beq
\delta \sum_{i=1}^N \sum_x  (-1)^x \cos (2k_Fx+\sqrt{2\pi} \phi^i) \, ,
\eeq{dimh}
which is incommensurate at non-zero magnetization, but can contribute
to radiatively generated commensurate terms as we discuss below
(see also \cite{CG2}).

Taking as an example $N=2$, one can show that the operator
\beq
\lambda' \sum_x (-1)^x
\cos (4k_Fx+\sqrt{4\pi} \phi^+)
\eeq{gapop}
is radiatively generated for non-zero magnetization, and it is
commensurate for $\langle M \rangle =1/2$.
This operator arises from a combined effect of the interchain coupling and
the dimerization along the chains and hence $\lambda' $ in (\ref{gapop}) is proportional
to $\Jp$,  $\delta$ and $\langle M \rangle$.
This operator has dimension $K$ and is hence relevant, implying the
existence of a plateau at $1/2$ of the saturation magnetization.

A related analysis was performed in \cite{Totsuka} but in the regime
where the interchain coupling $\Jp \rightarrow - \infty$, which
cannot be studied within the perturbative approach used  here.
In fact, in that case the existence of the plateau was associated with a
different operator of higher dimension, which could become relevant
only under certain conditions.

In the present case, the existence of the plateau is independent of
the values of the microscopic parameters, since the dimension of the operator
(\ref{gapop}) is always smaller than $2$. Furthermore, for a two-leg
zig-zag ladder there is only one way to introduce dimerization along
the legs and therefore in contrast to the ordinary ladders \cite{CG2}
cancellations do not occur.

An amusing observation which may also be of some relevance for
experiments is that one recovers the two-leg zig-zag ladder with
dimerized interchain coupling as the first-order effective
Hamiltonian for the two-leg ladder with dimerization along the chains
(see Fig.\ \ref{figdim2}) in the limit of $\delta \to 1$. Details are
given in an appendix. The main point is that dimerized interchain coupling
and dimerization along the legs break translational symmetry in
different ways: For the former case a translationally invariant
unit cell of the Hamiltonian contains two spins while in the
latter case it contains four. In view of the quantization condition
(\ref{condM}), this suggests that dimerization along the legs can
give rise to more plateaux than if only the interchain coupling
is dimerized.

A similar analysis as the one of this section can be performed for generic $N$.

\subsection{Doping with charge carriers}

Let us consider now a generalization of the system studied in
Section II by adding charge degrees of freedom. We start by describing the
Hamiltonian of interacting electrons in one dimension \cite{A2,ST}:
\beq
H = -{D \over 2} \sum_{x,\alpha} (c^{\dagger}_{x+1,\alpha} c_{x,\alpha}
+ H.c.) + U \sum_x c^{\dagger}_{x,+} c_{x,+} c^{\dagger}_{x,-} c_{x,-}
\eeq{HubHam}
where $c^{\dagger}$ and $c$ are electron creation and annihilation
operators and $\alpha = \pm$.
For positive $U$ and at half filling, the charge sector is massive
and the spin sector for large $U$ can be described by the Heisenberg
Hamiltonian \cite{A2}. Here, we will analyze the large scale
behaviour of two identical systems coupled in zig-zag, away from
half filling.
Since we are now keeping the $SU(2)$ symmetry of the spin sector
(neither magnetic field nor $XXZ$ anisotropy will be considered), the
continuum limit of the Hamiltonian at non-zero doping can be written as:
\beq
H = {1 \over 2} \int dx \left( v_c K_c (\partial_x \tilde{\phi}_c)^2
+ {v_c \over K_c} (\partial_x \phi_c)^2 \right)
+ H_{WZW}(g_s) - \lambda \int dx  \vec J_R \cdot \vec J_L \, ,
\eeq{IntWZW}
where the first term of the Hamiltonian stands for the charge sector
and the WZW term describes the spin sector.
The charge and spin density operators are given respectively by:
\beq
\rho (x) = j_R + j_L + const.~\sin(2 k_F x + \sqrt{2 \pi} \phi_c)
~\tr(g)
+ const.~\cos(4 k_F x + \sqrt{8 \pi} \phi_c)
\eeq{chargeOp}
and
\beq
\vec S(x)=\vec J_R + \vec J_L + const.~\sin(2 k_F x + \sqrt{2 \pi} \phi_c)
~\tr(\vec \sigma g) \, ,
\eeq{SpinOp}
where $j$ and $\vec J$ are the U(1) and $SU(2)$ currents of the charge
and spin sector respectively.
The Fermi momentum $k_F$ is now related to the charge sector and
we keep our spin sector at zero magnetization. It is known that
$K_c = 1$ for $U=0$ and $K_c = 1/2$ for $U = \infty$ \cite{FK}.

We again consider two copies of the system and study perturbatively
the zig-zag coupling between them. More precisely, we consider a
coupling term which involves both charge and spin densities
$$
H_{{\rm int}} = \sum_{x} \left( {U' \over 2} \rho^1_x (\rho^2_x + \rho^2_{x+1})
 + \Jp \vec S^1_x (\vec S^2_x + \vec S^2_{x+1}) \right).
$$
The presence of a further direct hopping term between the chains would require
a more detailed analysis. Indeed, first the Gaussian part must be
diagonalized including this direct hopping term and then, for example,
one can treat perturbatively terms like $H_{{\rm int}}$. This is beyond the
scope of the present article where we just concentrate on showing that
the presence of a weak coupling between Hubbard chains like in \cite{ST}
generalized to the zig-zag configuration can give rise to results
similar to those observed for normal ladders.

In the continuum description, this interaction term can be written in two
different pieces. We first have the current-current  interaction term given by
\beq
{U' \over 2} (j^1_R + j^1_L) (j^2_R + j^2_L)
+ \Jp (\vec J^1_R + \vec J^1_L) \cdot (\vec J^2_R + \vec J^2_L)
\eeq{nextInt}
The U(1) current-current term can be completely absorbed in the
Gaussian charge Hamiltonian by a rescaling of $v_c$ and $K_c$
\beq
K_c^{\pm} = (1 \pm {U' \over 2 \pi v_c})^{-1/2} K_c
\eeq{Kvals}
where the indices $\pm$ stand for the symmetric and antisymmetric fields
$(\phi_c^1 + \phi_c^2)/\sqrt{2}$ and $(\phi_c^1 - \phi_c^2)/\sqrt{2}$ respectively.
The $SU(2)$ current term is identical to the one of Section II.A
and we know then that it plays a r\^ole only for positive $\Jp$.

There is however another commensurate perturbation present
away from half filling which is given by
\bea
\cos k_F~\cos(\sqrt{4\pi}\phi_{c}^-)~\left( U'~const.~\tr(g^1)~\tr(g^2)
+ \Jp ~const.~\tr(\vec \sigma g^1) \cdot
\tr(\vec \sigma g^2) \right) \, .
\label{todos}
\eea
In the present case, the chirally asymmetric terms are irrelevant
in the RG sense.

In the weak interchain coupling limit we are considering, $U'$ is much smaller
 than $U$. Then $U'/(2 \pi v_c^-)< 1-K^2_c$ and this term is relevant
giving a mass to all the spin degrees of freedom as well as
to the antisymmetric charge field $\phi_c^-$. Thus, only the
field $\phi_c^+$ remains massless.

We see from (\ref{todos}) that also in this case an effective
perpendicular coupling
is generated, signaling the instability of the zig-zag
interchain coupling also against charge doping.

Since the perturbation terms in (\ref{todos}) are the same
as those that appear in the normal coupling studied in \cite{ST},
it is then natural to expect that the IR behaviour of
the charge density wave and superconducting correlation functions
will be the same in the present case.

\section{Numerical analysis of the two-leg case}

In this section we consider the $N=2$ version of (\ref{zzHam}) without
dimerization, \ie\ $\delta = 0$. It is then useful
to think of the two-leg zig-zag ladder as a single chain with next-nearest
neighbour interaction. So, for $N=2$ the Hamiltonian (\ref{zzHam}) can
be recast in the form (writing explicitly an $XXZ$ anisotropy $\Delta$)
\bea
H &=& \Jp \sum_{x=1}^L
\left\{ \Delta \Sz_x \Sz_{x+1} + {1 \over 2}
\left(S^{+}_x S^{-}_{x+1} + S^{-}_x S^{+}_{x+1} \right)\right\} \nn \\
&&+ J \sum_{x=1}^L
\left\{ \Delta \Sz_x \Sz_{x+2} + {1 \over 2}
\left(S^{+}_x S^{-}_{x+2} + S^{-}_x S^{+}_{x+2} \right)\right\}
  \label{latHamN2} \\
&& - h \sum_{x=1}^L \Sz_{x} \, . \nn
\eea
Here $L$ denotes the total volume of the system.
In this section we will always assume $J > 0$.
To avoid frustration also in the limit $\Jp \to 0$ where we
find two weakly coupled chains, we choose $L$ to be a multiple of 4.

In the formulation (\ref{latHamN2}) essentially all spatial symmetries
are manifestly implemented by a one-site translation $x \to x+1$.
So, one can use Fourier transforms to simplify the determination
of the spectrum. Since the magnetic field $h$ is coupled to
a conserved order parameter in (\ref{latHamN2}), we can relate all
quantities at a field $h$ to those at $h=0$ -- the results
to be reported below are all obtained from computations
with $h=0$.

There is already a number of exact diagonalization studies for the
two-leg $XXZ$ zig-zag ladder (or equivalently the Heisenberg chain
with next-nearest-neighbour interactions) in a magnetic field
\cite{ToHa1,ToHa2,ToHa3,SGMK,GFAMAK,GMK,UsSu,TNK}. However, there
are still some regions in the parameter space and aspects which
have not been studied in great detail, such that further numerical
investigations seem worthwhile. For a numerical analysis we concentrate
on the isotropic
point $\Delta = 1$. It turns out that groundstate momenta can be
incommensurate in the presence of the magnetic field. Such a feature
is interesting in its own right and reminds us of recent observations
made for the $S=1$ chain with biquadratic interaction \cite{FaLi,GJS},
but also gives rise to technical complications. This incommensurability is
one reason why the computations to be reported below needed substantially
more CPU time than analogous computations for conventional
ladders \cite{CHP}, even though we used an improved version of
the program employed loc.cit.

\begin{figure}[ht]
\psfig{figure=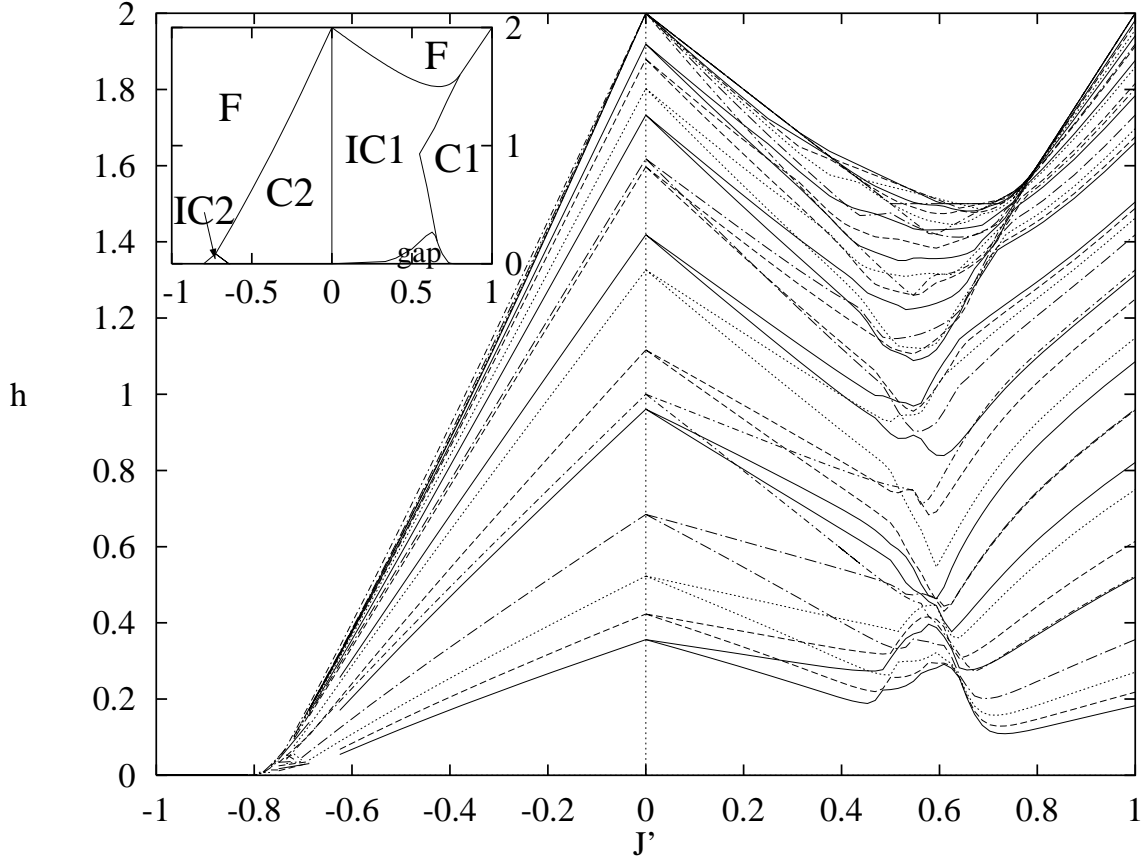,width=\columnwidth,angle=270}
\smallskip
\caption{
Magnetic phase diagram with the choice of normalization
$J + \abs{\Jp} = 1$. The lines are for $L=24$ (full),
$L=20$ (dashed). $L=16$ (dotted), $L=12$ (long dashed-dotted)
and $L=8$ (dashed-dotted). The inset shows a schematic version
which is discussed in the text.
\label{fig1}
}
\end{figure}

In order to scan the whole range of coupling constants we choose
a normalization such that $J + \abs{\Jp} = 1$. $\Jp = 0$ then
corresponds to two decoupled chains, $\Jp = 1$ to a single
antiferromagnetic chain and $\Jp = -1$ to a single ferromagnetic
chain. The resulting magnetic phase diagram is shown in Fig.~\ref{fig1}.
Here, the lines show the magnetic fields where the magnetization jumps
between two different values that are realized at a given system size.
The conclusions are schematically summarized in the inset of
Fig.~\ref{fig1}. Regions with a ferromagnetically polarized groundstate
are denoted by an `F'. The other regions will be explained in
the following discussion of our results.

Let us look first at the antiferromagnetic region $\Jp > 0$.
Here spins flip one
after the other. The groundstate momenta are in general
incommensurate (\ie\ not multiples of $\pi / 2$) in the region IC1
($0 < \Jp < 4 J$), though for the small lattice sizes accessible
to us, this does not show up in the region of small $\Jp$.
For $h=0$, a study of the static structure factor exhibited
a transition to incommensurate behaviour at $\Jp/J \approx 1.92075$
\cite{BGFPXZ}.

For $h>0$, the onset of incommensurability in the groundstate
momenta was determined in \cite{GFAMAK}, and \cite{FraRoe}
determined a transition in the groundstate using a different
criterion.
In the incommensurate region IC1 of Fig.~\ref{fig1}, the lines are irregular
which is mainly due to the lattice discretization of the momenta.
In the region C1, {\it i.e.}\
for $\Jp > 4 J$, all groundstate momenta are commensurate
and convergence with system size is good.

A gap \cite{CPKSR,WA} can be anticipated in Fig.~\ref{fig1} for
$J \le \Jp \le 2 J$, but apart from that there is no evidence
for any non-trivial plateaux. In fact, non-trivial plateaux
have not been observed at the various points which have been
studied over the past decade \cite{TNK,ToHa1,ToHa2,SGMK,UsSu}.
These observations can be nicely interpreted in terms of the
quantization condition (\ref{condM}). In view of the mapping
to a single chain (\ref{latHamN2}), one should substitute $N=1$
in (\ref{condM}). The gap then arises by spontaneous breaking
of this enhanced translational symmetry to $l=2$. Non-trivial
plateaux would then require $l > 2$, which at least in the
two-leg zig-zag ladder does not seem to be permitted.

The magnetization process on the ferromagnetic side has already been
looked at some time ago \cite{ToHa3}, though no definite
conclusion was reached due to problems of resolution. Indeed,
the ferromagnetic side is quite a bit different from the
antiferromagnetic one, even though plateaux do neither seem
to occur here. In the region C2 ($-3 J/2 \le \Jp < 0$),
spins flip in pairs and the groundstate momenta of those
states that participate in the magnetization process are commensurate
(actually $0$ or $\pm \pi$). The states with an odd number of spins pointing
up (or down) can have incommensurate momenta, but they do not participate
in the magnetization process.

At $\Jp = -4 J$ a transition to a completely polarized ferromagnet
takes place. The precise location of the transition point
can be attributed to the fact that at $\Jp = -4 J$ an exact $S=0$
groundstate, the uniformly distributed RVB state \cite{HKNN} can be
written down and is degenerate with the ferromagnetic state.
The intermediate region IC2 (\ie\ roughly $-4 J < \Jp < -3 J/2$)
is rather complicated: Here the number of spins flipping simultaneously
at a given system size changes with $\Jp / J$. Furthermore, groundstates
with incommensurate momenta (\ie\ momenta not an integer multiple of $\pi$)
do participate in the finite-size magnetization process.

\begin{figure}[ht]
\psfig{figure=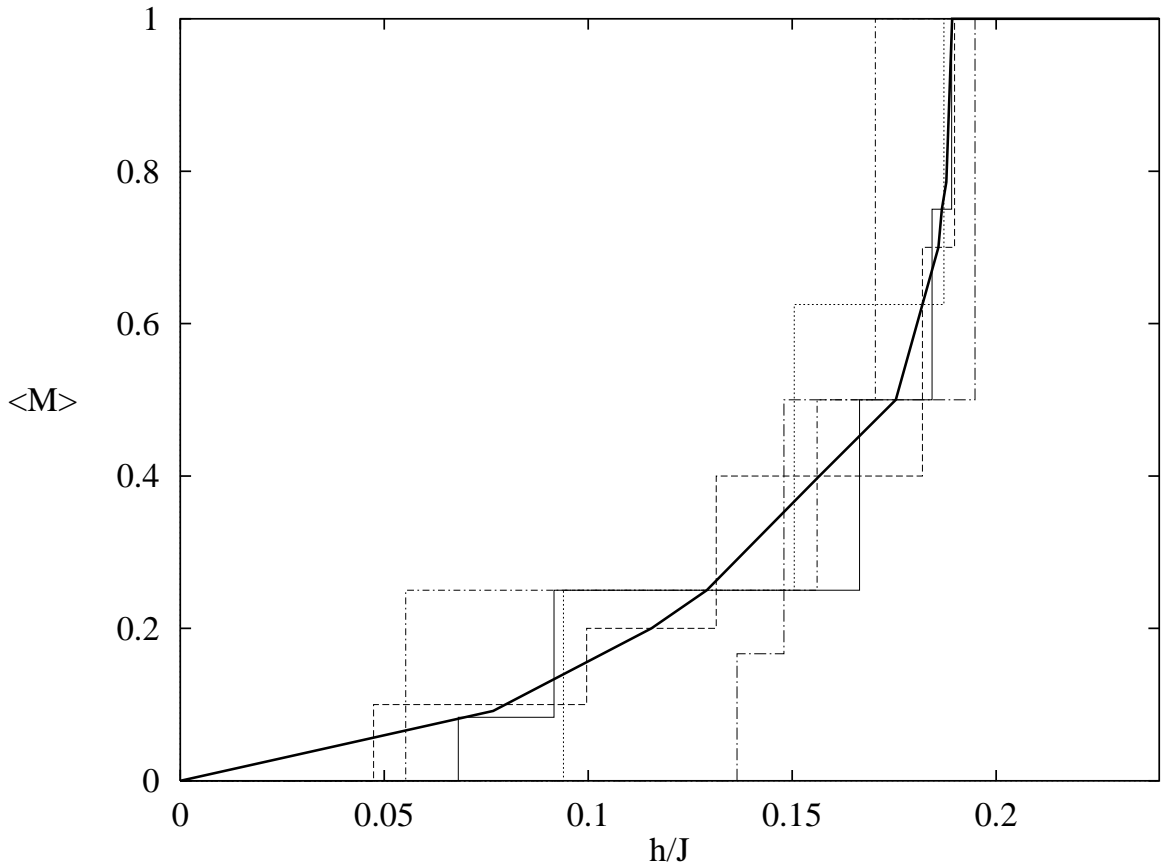,width=\columnwidth,angle=270}
\smallskip
\caption{
Magnetization curves at $\Jp = -3J$ with lines for $L=24$ (full),
$L=20$ (dashed). $L=16$ (dotted), $L=12$ (long dashed-dotted)
and $L=8$ (dashed-dotted). The bold line is a sketch of the
form expected in the thermodynamic limit.
\label{fig2}
}
\end{figure}

The magnetization process in the region IC2 is illustrated in
Fig.~\ref{fig2} by the finite-size magnetization curves at $\Jp = -3J$.
Here, the number of spins that flip simultaneously in a finite-size system
varies between $1$ and $3$ (the number of spins flipping simultaneously
increases as one approaches $\Jp = -4J$).
Strong non-monotonic finite-size effects can be observed in particular
at the smaller system sizes. Nevertheless, a reasonable approximation
to the limit $L \to \infty$ seems to be obtained applying the
procedure of \cite{BoFi} to $L=20$ and $L=24$, \ie\ by connecting
the midpoints of the steps in the finite-size magnetization curves.
This yields the bold line in Fig.~\ref{fig2}. The behaviour
at small fields is somewhat speculative since the finite-size
gaps are non-monotonic such that it is not possible to extrapolate them.
We have therefore simply assumed a vanishing gap in the limit $L\to \infty$.

The region close to saturation in Fig.~\ref{fig2} is on comparably
safe grounds. In fact, we have data for up to $L=48$ which has
been taken into account for the bold curve. In the case $\Jp = -3J$,
this data yields no evidence for more than three spins flipping
simultaneously at the transition $\langle M \rangle \to 1$. Therefore
we are confident that the magnetization
curve becomes smooth in the thermodynamic limit (though very
steep at the transition $\langle M \rangle \to 1$). In
particular, we think that our intermediate phase IC2 is different
from the metamagnetic one observed in \cite{GMK} in a different
parameter region, even though at first sight they bear
some resemblance. The fact that we are not aware of any evidence
for incommensurate momenta arising in the metamagnetic
phase \cite{GMK} also suggests that there are two distinct
phases preceding a transition to a ferromagnet in different
parameter regions.

We would like to conclude the present discussion with a few remarks
on the spin-gap. Although it has already been investigated with high
accuracy for $\Jp > 0$ applying the density matrix renormalization
group to large system sizes \cite{CPKSR,WA,Sor}, there are still
some interesting problems. Firstly, only two of these investigations
\cite{CPKSR,WA} deal with the incommensurate region and there the
results differ by more than 10\%. Secondly, no such investigations were
performed for $\Jp < 0$. In these two regions, $S=1$ excitations
(corresponding to the lowest lines in Fig.\ \ref{fig1} for $\Jp > 0$)
have an incommensurate momentum at the minimum of their dispersion.
Since the system size limits the resolution in momentum-space, one has
additional non-monotonic finite-size effects. We have tried to overcome
this with an interpolation using a Fourier transform of the dispersion.
However, even/odd momentum effects (which are characteristic for scattering
states) obscure the true incommensurate minimum and preclude
such an analysis. According to preliminary investigations,
such an approach is applicable to $S=1/2$ excitations (the fundamental
spinon-type excitations) not only in the commensurate region \cite{Sor}
but also in the incommensurate region.

The question of a spin-gap is particularly interesting for $\Jp < 0$.
In Section II.A we have confirmed the conclusion of \cite{NGE} that
a one-loop RG analysis suggests a non-zero spin-gap for small $\Jp < 0$.
However, for $\Jp < 0$ there is no region where a gap in the magnetic
excitations is obvious from our numerical data. The data for the
$S^z = 2$ excitations (corresponding to the lowest lines for
$\Jp < 0$ in Fig.\ \ref{fig1}) actually fits well to the form
\beq
\Gap(L) \sim {1 \over L} \, .
\eeq{GapConfInf}
Such a form is characteristic for a conformally invariant and thus gapless
situation (see e.g.\ \cite{Cardy}). For this reason we have not drawn
a gap in the schematic inset of Fig.\ \ref{fig1} for $\Jp < 0$. Nevertheless,
it remains a challenge to find numerical evidence for a spin-gap in some
magnetic excitations and look for possible massless sectors which should
be organized into $\widehat{su(2)}_1$ representations.

\section{The transition to saturation in the two-leg ladder}

\subsection{The antiferromagnetic side}

Now let us look at the transition to saturation in the lattice-version
of the $N=2$-leg zig-zag ladder.

First we recall the computation of the transition field $h_{{\rm uc}}$
\cite{GMK}. One can immediately write down the energy of a fully polarized
(ferromagnetic) state (with $h=0$)
\beq
E_{{\rm sat.}} = {L \over 4} \Delta \left(\Jp + J\right) \, .
\eeq{Esat}
Using a Fourier transform, also the excitation energy of a
single spin-wave above this fully polarized state is readily
computed as
\beq
{\cal E}_{1s}(p) = - \Delta \left(\Jp + J\right) + \Jp \cos{p} + J \cos(2p)
   \, .
\eeq{E1s}
To determine the critical magnetic field $h_{{\rm uc}}$ associated
to the transition $\langle M \rangle \to 1$, we have to minimize
(\ref{E1s}). One finds the minimum at
\beq
p_{{\min.}} = \cases{ \pi - \cos^{-1} \left({\Jp \over 4 J}\right) &
                    for $0 \le \Jp \le 4 J$, \cr
              \pi      & for $\Jp \ge 4 J$. \cr}
\eeq{Pmin}
An immediate consequence of this simple result is that the groundstate
of the $S^z = L-1$ sector has an incommensurate momentum
for $0 < \Jp < 4 J$. This is presumably the simplest example of
how the two competing interactions in (\ref{latHamN2}) lead to the
phase with incommensurate groundstate momenta which we discussed
in the previous section.

Insertion of (\ref{Pmin}) into (\ref{E1s}) directly leads to the
upper critical field
\beq
h_{{\rm uc}} = \cases{ \Delta \Jp + (\Delta + 1) J + {\Jp ^2 \over 8 J}
                        & for $0 \le \Jp \le 4 J$, \cr
                       (\Delta  + 1) \Jp + (\Delta - 1) J
                        & for $\Jp \ge 4 J$. \cr}
\eeq{huc}
\indent
Now we will go beyond \cite{GMK} and discuss the nature of the transition
$\langle M \rangle \to 1$. To this end, it is useful to look at
two-spinwave excitations \cite{HoPa}. After a
Fourier transformation, one arrives at a matrix problem in the
distance of the two flipped spins. We omit the explicit expressions
for the matrices that we have used to study the behaviour of the
magnetization curve close to saturation. This is equivalent to studying the
finite-size behaviour of the two-spinwave groundstate energy
${\cal E}_{2s}$ at $h=h_{{\rm uc}}$, since
$1 - \langle M \rangle = 4/L$ and ${\cal E}_{2s} = h - h_{{\rm uc}}$.

For the two-leg zig-zag ladder, the eigenvalue problem in the two-spinwave
subspace can be interpreted as a fourth-order difference equation with
suitable boundary conditions and thus can in principle be solved
by a further Fourier transformation. For our purposes it turned out to
be sufficient to fix the center-of-mass momentum $p$ and then perform a
numerical diagonalization for system sizes up to $L \approx 150$. We have
looked at $\Delta = 1$ and the following values of
the coupling constants: $\Jp = 2 J$, $\Jp = 4 J$,
$\Jp = 4.1 J$ and $\Jp = 4 \cos(\pi / 12) J$. These values were
selected since for them (\ref{Pmin}) can be realized exactly for suitable
choices of the system size $L$ and one can thus avoid further
non-monotonic finite-size effects which could arise from the lattice
discretization in momentum space. In all cases that we have studied,
the minimum energy of the two-spinwave excitation was found in the
sector with center-of-mass momentum $p=0$.

The numerical diagonalization determines the critical field
associated to the transition from two flipped spins to one
flipped spin. For $\Jp \ne 4 J$, this critical field is compatible with the
universal DN-PT behaviour \cite{DzNe,PoTa}
\beq
h_{{\rm uc}} - h \sim (1 - \langle M \rangle)^2 \, .
\eeq{DNPT}
Just at $\Jp = 4 J$ we find a different behaviour which is
much better described by
\footnote{The exponent $4$ has also been observed in \cite{SGMK}
at $\Jp/J = 4$.}
\beq
h_{{\rm uc}} - h \sim (1 - \langle M \rangle)^4 \, .
\eeq{nDNPT}
\indent

To understand why just the point $\Jp = 4 J$ leads to a different
exponent, it is instructive to look at
the behaviour of ${\cal E}_{1s}$ near the minimum $p_{{\min.}}$.
{}From (\ref{E1s}) and (\ref{Pmin}) we find
\beq
{\cal E}_{1s}(p_{{\min.}}+x) - {\cal E}_{1s}(p_{{\min.}})
 = \cases{ {(4 J - \Jp) (4 J + \Jp) \over 8 J} x^2 + {\cal O}(x^3)
                 & for $\Jp < 4 J$, \cr
           {J \over 2} x^4 + {\cal O}(x^6)
                 & for $\Jp = 4 J$, \cr
           \left({\Jp \over 2} - 2 J\right) x^2 + {\cal O}(x^4)
                 & for $\Jp > 4 J$. \cr }
\eeq{dispPmin}
We see that the dispersion is quadratic everywhere except for
the point $\Jp = 4 J$ where it is quartic. These exponents
in the one-spinwave dispersion are in one-to-one correspondence with
those in (\ref{DNPT}) and (\ref{nDNPT}), respectively. So, the
behaviour of the magnetization for $\langle M \rangle \to 1$ can
be explained by the band structure.
Note that (\ref{dispPmin}) is independent of $\Delta$, and so should
be the exponents corresponding to (\ref{DNPT}) and (\ref{nDNPT}),
respectively (as long as $\Delta > 0$).

\subsection{The ferromagnetic side}

Now we turn to the transition to saturation on the ferromagnetic side
$\Jp < 0$. Since we know from the numerical investigation that spins flip in
pairs (as long as $\abs{\Jp}$ is not too large), we have to solve
a non-trivial matrix problem already to determine the
critical field $h_{{\rm uc}}$ in the ferromagnetic regime $\Jp < 0$.
For $\abs{\Jp} < 2.5 J$, the minimum of the dispersion of the
two-spinon state is now located at $p = \pi$. In the region
$\Jp < -2.66 J$, the minimum moves away form $p = \pi$,
{\it i.e.}\ it becomes incommensurate. In the following we
restrict ourselves to the region of sufficiently small $\abs{\Jp}$
such that spins flip in pairs and only commensurate groundstates
participate in the magnetization process for $\langle M \rangle \to 1$.

Some values of transition fields $h_{{\rm uc}}$ are listed
in Table~\ref{tabsat}. These have actually been obtained
on chains with a few hundred sites. Nevertheless,
all given digits should be those of the thermodynamic limit.

To check the asymptotic behaviour of the magnetization curve,
one further needs at least
the transition field $h_{4 \to 2}$ at which two spins flip
to reduce the number of spins deviating from the ferromagnetic
state from four to two. Table~\ref{tabsat} lists transition
fields $h_{4 \to 2}$ computed numerically on chains with
$36 \le L \le 72$. Since $1 - \langle M \rangle \sim 1/L$,
we can then test for the universal behaviour (\ref{DNPT})
by forming the expression
\beq
L \sqrt{h_{{\rm uc}} - h_{4 \to 2}} \, ,
\eeq{testDNPT}
which should converge to a constant if (\ref{DNPT}) is satisfied.
Indeed, the expression (\ref{testDNPT}) appears to converge to
a constant with incresing $L$ for all cases listed in Table~\ref{tabsat}.
So, we find no counterevidence against the universal DN-PT
behaviour on the ferromagnetic side either, although we have
no compelling argument in its favour.

\begin{table}[ht]
\begin{tabular}{c|cccccccccc}
$\Jp/J$ & $0$     & $-3/13$ & $-1/3$  & $-5/11$ & $-3/5$  & $-7/9$
                          & $-1$ & $-9/7$  & $-5/3$ & $-11/5$ \\ \tableline
$L$     & \multispan{10} $h_{4 \to 2} / J$ \\ \tableline
$36$    & $1.96595$ & $1.77368$ & $1.69888$ & $1.60799$ & $1.50460$ & $1.38523$
                          & $1.24387$ & $1.07095$ & $0.85064$ & $0.55455$ \\
$40$    & $1.97272$ & $1.78215$ & $1.70116$ & $1.60981$ & $1.50630$ & $1.38682$
                          & $1.24527$ & $1.07208$ & $0.85145$ & $0.55495$ \\
$44$    & $1.97766$ & $1.78427$ & $1.70263$ & $1.61109$ & $1.50753$ & $1.38796$
                          & $1.24624$ & $1.07287$ & $0.85201$ & $0.55522$ \\
$48$    & $1.98137$ & $1.78568$ & $1.70366$ & $1.61204$ & $1.50844$ & $1.38879$
                          & $1.24695$ & $1.07343$ & $0.85241$ & $0.55542$ \\
$52$    & $1.98423$ & $1.78666$ & $1.70441$ & $1.61277$ & $1.50914$ & $1.38941$
                          & $1.24748$ & $1.07385$ & $0.85271$ & $0.55556$ \\
$56$    & $1.98648$ & $1.78736$ & $1.70499$ & $1.61335$ & $1.50968$ & $1.38988$
                          & $1.24788$ & $1.07418$ & $0.85294$ & $0.55567$ \\
$60$    & $1.98828$ & $1.78788$ & $1.70546$ & $1.61381$ & $1.51010$ & $1.39025$
                          & $1.24819$ & $1.07443$ & $0.85312$ & $0.55576$ \\
$64$    & $1.98974$ & $1.78828$ & $1.70583$ & $1.61418$ & $1.51044$ & $1.39054$
                          & $1.24844$ & $1.07463$ & $0.85326$ & $0.55582$ \\
$68$    & $1.99094$ & $1.78859$ & $1.70615$ & $1.61448$ & $1.51071$ & $1.39078$
                          & $1.24864$ & $1.07479$ & $0.85338$ & $0.55588$ \\
$72$    & $1.99195$ & $1.78885$ & $1.70641$ & $1.61473$ & $1.51093$ & $1.39097$
                          & $1.24880$ & $1.07492$ & $0.85347$ & $0.55592$ \\
        \tableline
$h_{\rm uc}/J$ &
          $2$     &$1.79086$&$1.70833$&$1.61648$&$1.51250$&$1.39236$
                      & $1.25000$&$1.07589$&$0.85417$&$0.55625$ \\
\end{tabular}
\caption{Values of $h_{4 \to 2}$ at which a transition from four to
two spins deviating from the ferromagnetic state takes place at
a given size $L$. The last row contains auxiliary data, {\it i.e.}\
values of $h_{\rm uc}$ which were estimated using much larger systems.
}
\label{tabsat}
\end{table}

We expect that the transition $\langle M \rangle \to 1$ remains
second order if we increase $\abs{\Jp}$ further into the regime
where more than two spins flip at the same time (this is in
contrast to the metamagnetic transition studied in \cite{GMK}).
At least in this region the very steep behaviour of the magnetization
curve (see Fig.\ \ref{fig2}) could mean a breakdown of the
DN-PT universal behaviour (\ref{DNPT}). However, due to the flipping
of several spins at the same time, we are not able to investigate
the transition to saturation in detail numerically in this region.

\section{The three-leg case}

Now we proceed with a discussion of $N=3$-leg zig-zag ladders
on the lattice, though in less detail than for the two-leg ladder. For
$N=3$ there is a large number of possible boundary conditions
for the coupling between the chains. In particular for periodic
boundary conditions (PBC), there are several possibilities
already for $N=3$, of which none is naturally singled out in
the case of zig-zag coupling. We will discuss three types of
PBC in addition to open boundary conditions along the rungs.
To elucidate these different types of boundary conditions it
may be useful to consider the case $\Jp = J$ where the three-leg
zig-zag ladder can be considered as a strip of the triangular
lattice Heisenberg antiferromagnet.
The triangular lattice has three sublattices and it is possible
to consider boundary conditions that do or do not respect this
sublattice structure. We will call the PBC that respect this
sublattice structure at $\Jp = J$ `PBC of type A', while those
that identify different sublattices `PBC of type B and C'.

The $N=3$-leg zig-zag ladder
with PBC of type B and C can be rewritten as a single Heisenberg chain with
interactions up to distances of three, respective four sites:
\bea
H &=& \Jp \sum_{n=1}^{3 L}
\left\{ \Delta \Sz_n \Sz_{n+1} + {1 \over 2}
\left(S^{+}_n S^{-}_{n+1} + S^{-}_n S^{+}_{n+1} \right)
+ \Delta \Sz_n \Sz_{n+r} + {1 \over 2}
\left(S^{+}_n S^{-}_{n+r} + S^{-}_n S^{+}_{n+r} \right)\right\} \nn \\
&&+ J \sum_{n=1}^{3 L}
\left\{ \Delta \Sz_n \Sz_{n+3} + {1 \over 2}
\left(S^{+}_n S^{-}_{n+3} + S^{-}_n S^{+}_{n+3} \right)\right\}
- h \sum_{n=1}^{3 L} \Sz_{n} \, .
  \label{latHamN3pbcF}
\eea
Here the second interaction term goes over a distance $r=2$ for
PBC of type B, while PBC of type C are characterized by $r=4$.

Due to (\ref{latHamN3pbcF}), the three-leg zig-zag ladder with
PBC of type B can be considered as
the natural counterpart of the $N=2$ zig-zag ladder where the
Hamiltonian can be written in the form (\ref{latHamN2}).

In contrast to the preceding discussion of the two-leg case we
will from now on again denote the length of each chain by $L$,
\ie\ the total number of spins is then $3 L$. Accordingly, we
introduce a momentum $p$ by a one-site translation along one
of the three chains.

In the following we will study each of these four boundary conditions
in turn.

\subsection{The saturation field}

First, we compute the field $h_{{\rm uc}}$ at which the transition
to saturation takes place. This is not only useful to check the
numerical computation to be reported in the next subsection,
but like in the case $N=2$ also serves as a guide where
groundstates with incommensurate momenta can participate in the
magnetization process.

\medskip
\leftline{{\it I) PBC of type A:}}

A simple computation yields the excitation energy of one flipped
spin above a ferromagnetic background as
\beq
{\cal E}_{1s,\pm}(p) =
-2 \Delta \Jp - {\Jp \over 2}
       \left(1 + \cos(p) \pm \sin(p) \sqrt{3}\right)
-\Delta J + J \cos(p) \, .
\eeq{E1sNFpbc}
The minimum of this dispersion is located at
\beq
\tan\left(p_{\min.,\pm}\right) = \pm {\Jp \sqrt{3} \over \Jp - 2 J} \, ,
\eeq{pminNFpbc}
which is generically incommensurate. This then leads to
the transition field
\beq
h_{{\rm uc}} = - {\cal E}_{1s,\pm}(p_{\min.,\pm}) =
\Delta \left (2 \Jp + J\right ) + {\Jp \over 2}
+ \sqrt{\Jp^2 - \Jp J +J^2} \, .
\eeq{hucNFpbc}

\medskip
\leftline{{\it II) Open boundary conditions:}}

The three-leg zig-zag ladder with open boundary conditions is
shown in Fig.\ \ref{figdim1}. Here we consider only the case
without dimerization ($\delta = 0$).

The dispersion of the gap corresponding to a single flipped spin
above the ferromagnetic background is given by
\beq
{\cal E}_{1s}(p) = - \Delta J + J \cos\left(p\right)
          -\Jp \left({3 \over 2} \Delta + {1 \over 2}
          \sqrt{\Delta^2 + 8 \cos\left({p \over 2}\right)^2} \right) \, .
\eeq{E1sOBC}
For $\Delta J < \Jp < J \sqrt{\Delta^2 + 8}$, this dispersion
has an incommensurate minimum at
\beq
\cos\left({p_{\min.} \over 2}\right)
= {1 \over 4} {\sqrt{2 \Jp^2 - 2 \Delta^2 J^2} \over J} \, .
\eeq{pminOBC}
This then leads to an upper critical field
\beq
h_{\rm uc} = -{\cal E}_{1s}(p_{\min.}) =
\left({\Delta \over 2}  + 1\right)^2 J + {\Jp^2 \over 4 J}
         + {3 \Jp \Delta \over 2} \, .
\eeq{hucOBC}

\medskip
\leftline{{\it III) PBC of type B and C:}}

Using the representation (\ref{latHamN3pbcF}) one easily finds the
dispersion of a single
flipped spin above the ferromagnetic background
\beq
{\cal E}_{1s}(p) = - \Delta (2 \Jp + J) + J \cos\left(p\right)
       + \Jp \left\{ \cos\left({r p \over 3}\right)
           + \cos\left({p \over 3}\right) \right\} \, .
\eeq{E1sFpbc}
Since the PBC of type B ($r=2$) are simpler to discuss analytically,
we will now concentrate on this case. The dispersion
(\ref{E1sFpbc}) then always has an incommensurate minimum at
\beq
\cos\left({p_{\min.} \over 3}\right)
= {\sqrt{\Jp^2-3 J \Jp + 9 J^2} -\Jp \over 6 J} \, .
\eeq{pminFpbc}
{}From this we find an upper critical field for PBC of type B
\beq
h_{\rm uc} = -{\cal E}_{1s}(p_{\min.}) =
\Delta \left (2 \Jp +J\right )
+ {\Jp \left (9 J \Jp -2 \Jp^2 +27 J^2 \right ) \over 54 J^2 }
+ {\left (\Jp^2 -3 J \Jp +9 J^2 \right )^{3 \over 2} \over 27 J^2 } \, .
\eeq{hucFpbc}

\subsection{Magnetization curves obtained by exact diagonalization}

Now we present one example of a magnetization curve for each of
the four boundary conditions introduced above. In all cases we
consider the $SU(2)$ symmetric situation $\Delta = 1$ and set
$\Jp/J = 3/2$.

\medskip
\leftline{{\it I) PBC of type A:}}

\begin{figure}[ht]
\psfig{figure=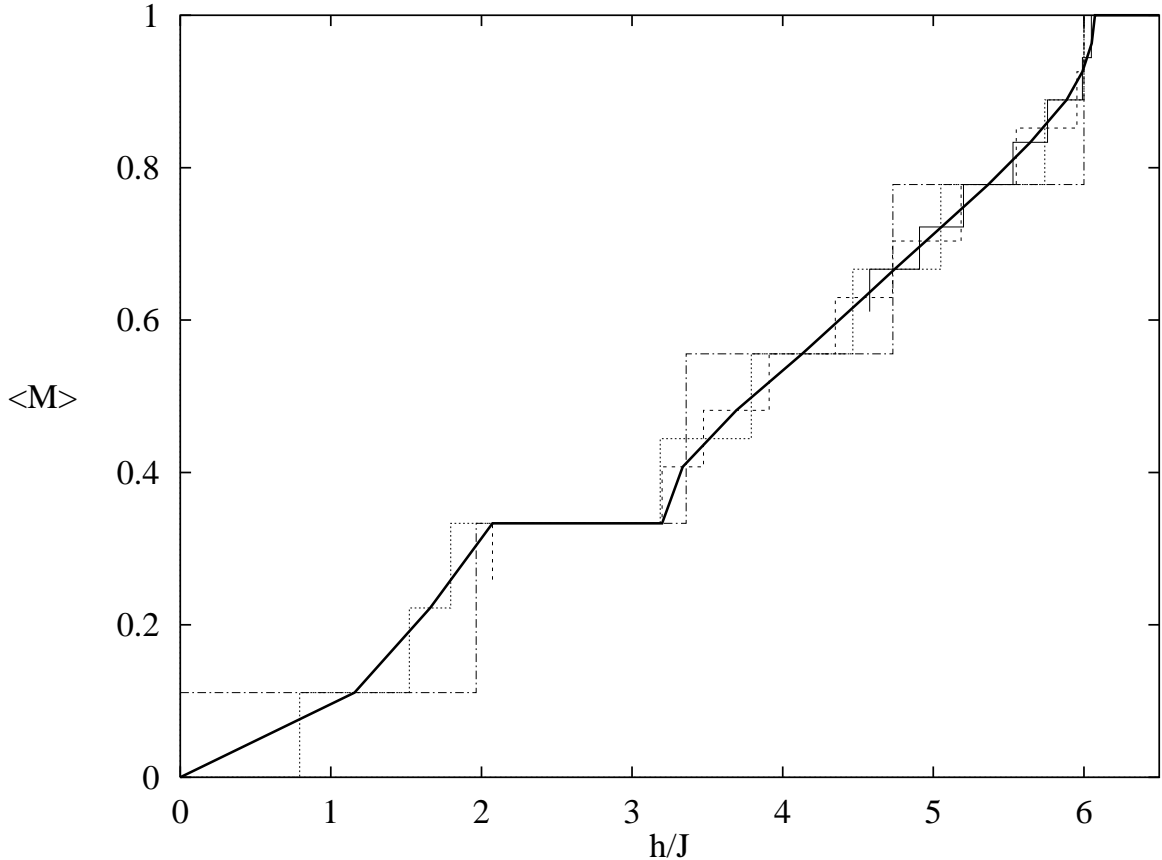,width=\columnwidth,angle=270}
\smallskip
\caption{
Magnetization curves of an $N=3$ zig-zag ladder with PBC of type A
at $\Jp = 3 J / 2$.
The lines are for $L=12$ (full), $L=9$ (dashed). $L=6$ (dotted)
and $L=3$ (dashed-dotted). The bold line is an extrapolation
to the thermodynamic limit.
\label{fig3}
}
\end{figure}

An example of a magnetization curve for $N=3$ with PBC of type A
is shown in Fig.~\ref{fig3}.
For $J=\Jp$, this geometry is one of the possible approximations
to a two-dimensional triangular lattice antiferromagnet and a
study of precisely this geometry \cite{MiNi,twodim} exhibits a clear plateau
with $\langle M \rangle = 1/3$. For $J = 0$, one recovers an `ordinary'
rectangular spin ladder with periodic boundary conditions which
then has equal coupling constants. The latter has already been
investigated in some detail \cite{CHP,CHP2,TLPRS} and evidence was
found for a small plateau with  $\langle M \rangle = 1/3$ (there is
further the possibility of a tiny spin-gap at $h=0$ \cite{KaTa,CHP2}).

In Fig.\ \ref{fig3} we chose $\Jp = 3 J / 2$ in order to
study a new example. One can see from (\ref{pminNFpbc}) that
this gives rise to an incommensurability for $\langle M \rangle \to 1$
(also at other values of the magnetization there is evidence
that this choice of parameters lies in an incommensurate phase).
Both at $J=0$ and $\Jp =0$, one
should choose even $L$ to avoid frustration by the periodic
boundary conditions along the chains, while for $\Jp = J$
frustration is avoided if $L$ is chosen to be a multiple
of $3$. To avoid both effects, $L$ should be chosen as a
multiple of $6$. However, this would considerably restrict
the system sizes accessible to us. We therefore require only
divisibility by three (which seems to be the more important one
if the coupling constants are of a similar magnitude).

The upper critical field for the choice of parameters in Fig.\ \ref{fig3}
is evaluated from (\ref{hucNFpbc}) as $h_{\rm uc}/J =
19/4 + \sqrt{7}/2 \approx 6.0729$.

Like for $N=2$ in Fig.\ \ref{fig2},
we applied the extrapolation procedure of \cite{BoFi}
to the largest available system sizes in order
to obtain the estimate for the magnetization curve in the
thermodynamic limit which is shown by the bold line in Fig.\ \ref{fig3}.

In this extrapolated magnetization curve we have only drawn an
$\langle M \rangle = 1/3$ plateau but no further ones. This
is motivated by the criterion (\ref{GapConfInf}):
If we compare $\Gap(6) \approx 0.793 J$ with $\Gap(4) \approx 1.274 J$
(the latter is not shown in Fig.\ \ref{fig3})
we see that the decrease is faster than (\ref{GapConfInf}). These are
only two data points which probably do not lie in the asymptotic regime
(in particular non-monotonic finite-size corrections may still be
important). It should also be noted that the weak-coupling analysis of
Section II.A predicts a spin-gap. However, Fig.~\ref{fig3}
corresponds to a rather large value $\Jp$. In any case, according
to our numerical data, a spin-gap seems to be at least very small
if present at all. We have therefore not drawn
an $\langle M \rangle = 0$ plateau (corresponding to a spin-gap)
in Fig.\ \ref{fig3}, though the speculative nature of
the extrapolation in the region of small fields should be kept in mind.

For non-zero magnetizations, we use the following generalization of
(\ref{GapConfInf}) as the criterion for a vanishing plateau width
(compare also \cite{SaTa}):
\beq
h_{c_2}(L) - h_{c_1}(L) \sim {1 \over L} \, .
\eeq{ConfInf}
At $\langle M \rangle = 2/3$ we find $h_{c_2} - h_{c_1}
\approx 0.993 J$, $0.584 J$, $0.529 J$ and $0.331 J$ for $L=4$, $6$,
$8$ and $12$, respectively.
Although we cannot entirely rule out a tiny $\langle M \rangle = 2/3$
plateau on the basis of this data, it appears to be quite unlikely.
We have therefore neither drawn a plateau at $\langle M \rangle = 2/3$
in Fig.\ \ref{fig3}.
In any case, our main result for PBC of type A is that one can observe
a clear plateau with $\langle M \rangle = 1/3$.

\medskip
\leftline{{\it II) Open boundary conditions:}}

The limit $J \to 0$ of an $N$-leg zig-zag ladder with open boundary
conditions gives rise to the recently introduced
`diagonal ladders' \cite{SMWSD}. In particular, for $N=3$ with
open boundary conditions one recovers the necklace ladder at
$J = 0$. This necklace ladder is very similar to the $S=(1,{1\over 2})$
ferrimagnetic chain which is known to exhibit a plateau with
$\langle M \rangle = 1/3$ \cite{Kuramoto,MSBMY}.

\begin{figure}[ht]
\psfig{figure=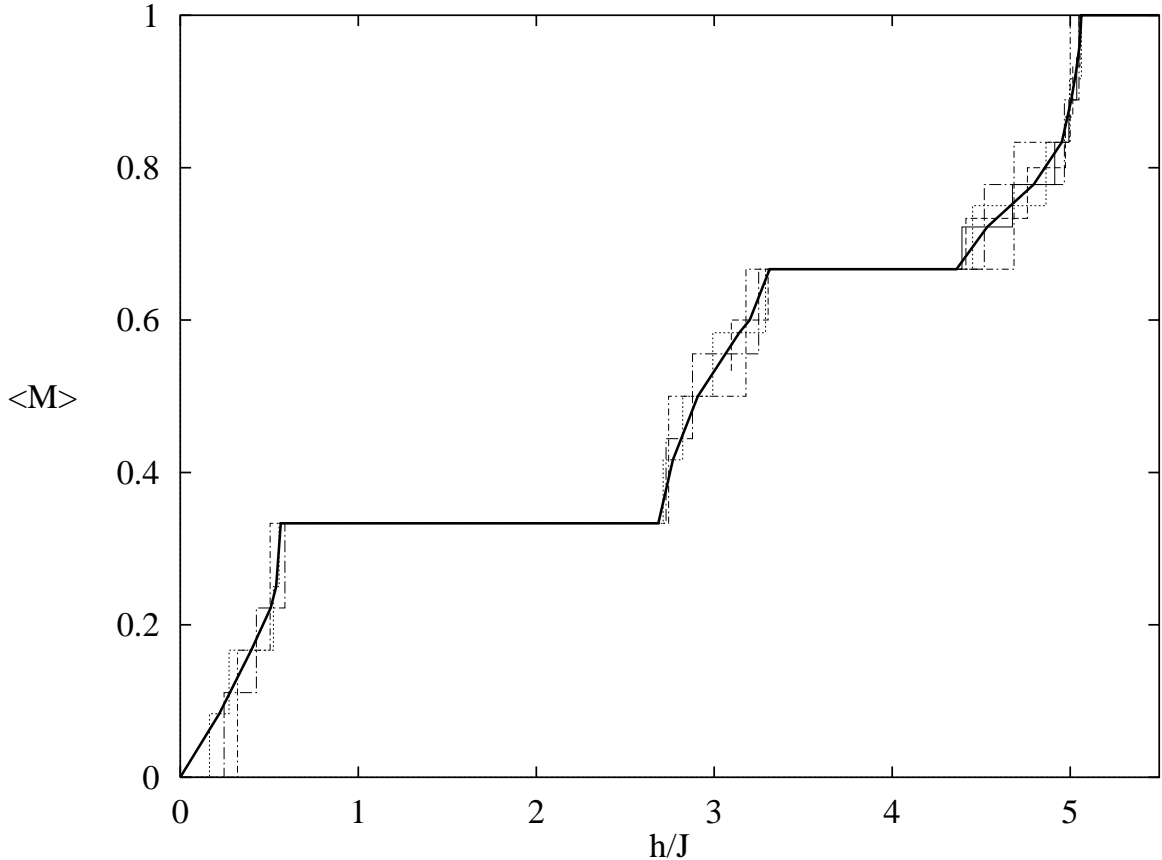,width=\columnwidth,angle=270}
\smallskip
\caption{
Magnetization curves of an $N=3$ zig-zag ladder with
open boundary conditions along the rungs and $\Jp = 3 J / 2$.
The lines are for $L=12$ (full), $L=10$ (dashed). $L=8$ (dotted),
$L=6$ (long dashed-dotted) and $L=4$ (dashed-dotted).
The bold line is an extrapolation to the thermodynamic limit.
\label{fig4}
}
\end{figure}

This strong-coupling limit $J \to 0$ suggests to choose $L$
divisible by two, as does the weak-coupling limit $\Jp \to 0$.
On the other hand, at $\Jp = J$ the $N=3$ zig-zag ladder with
open boundary conditions can be again considered as a strip
of a triangular lattice which would suggest that $L$ should be
chosen a multiple of three. For the choice of parameters
$\Jp/J = 3/2$ shown in Fig.\ \ref{fig4} divisibility by three turns out
to be not important and we impose only divisibility by two.

The choice of parameters corresponding to Fig.\ \ref{fig4}
($\Delta = 1$, $\Jp = 3 J/2$) appears to lie entirely in
an incommensurate phase. At least at the transition
$\langle M \rangle \to 1$, the momentum given by (\ref{pminOBC})
is clearly incommensurate. The upper critical field is then
found from (\ref{hucOBC}) as $h_{\rm uc} = {81 \over 16} J = 5.0625 J$.

The finite-size magnetization curves in Fig.\ \ref{fig4} exhibit
the expected plateau at $\langle M \rangle = 1/3$. In addition,
there is clear evidence for a further plateau at $\langle M \rangle = 2/3$.
The boundaries of these two plateaux were extrapolated applying either
a Shanks transform (which is the $\alpha= 0$ special case of
the vanden Broeck--Schwartz algorithm -- see e.g.\ \cite{HeSchue})
to their values $h_{c_i}(L)$ at the available system sizes $L$,
or (in the case of the upper boundary of the $\langle M \rangle = 1/3$
plateau) by fitting to
\beq
h_{c_i}(L) = h_{c_i}(\infty) + {a \over L} \, .
\eeq{1overLcorr}
This heuristic formula is motivated by (\ref{ConfInf}), since
eq.\ (\ref{1overLcorr}) can be expected to yield coincident
plateau boundaries if there is actually no plateau. In general,
(\ref{1overLcorr}) will work well if it is applied to system sizes
substantially below the correlation length and underestimate the width
of a plateau otherwise. Here, it gives a small correction to the value
for $h_{c_i}$ obtained for the largest $L$, as does the Shanks
extrapolation at the other plateau boundaries.

The finite-size spin-gap
for $L=4$, $6$ and $8$ is nicely fitted by (\ref{GapConfInf}).
This indicates that there is no spin-gap at $h=0$ and we have therefore not
drawn an $\langle M \rangle = 0$ plateau in the extrapolated magnetization
curve of Fig.\ \ref{fig4}.
Recently it was shown that a related three-leg ladder at zero field
is also massless and gives rise to a $c=2$ ($\widehat{su(2)}_2 \times$
Ising) conformal field theory \cite{ALN}. However there
the chirally asymmetric perturbation was eliminated by a particular
choice of coupling in order to permit an analytic treatment.
In the present case the central charge might therefore
be smaller than two.

The bold line in Fig.\ \ref{fig4} is an extrapolation taking into
account the foregoing discussion of plateau boundaries.
Between the plateaux it has been obtained by the same procedure
as used in Figs.\ \ref{fig2} and \ref{fig3}.

\medskip
\leftline{{\it III) PBC of type B and C:}}

\begin{figure}[ht]
\psfig{figure=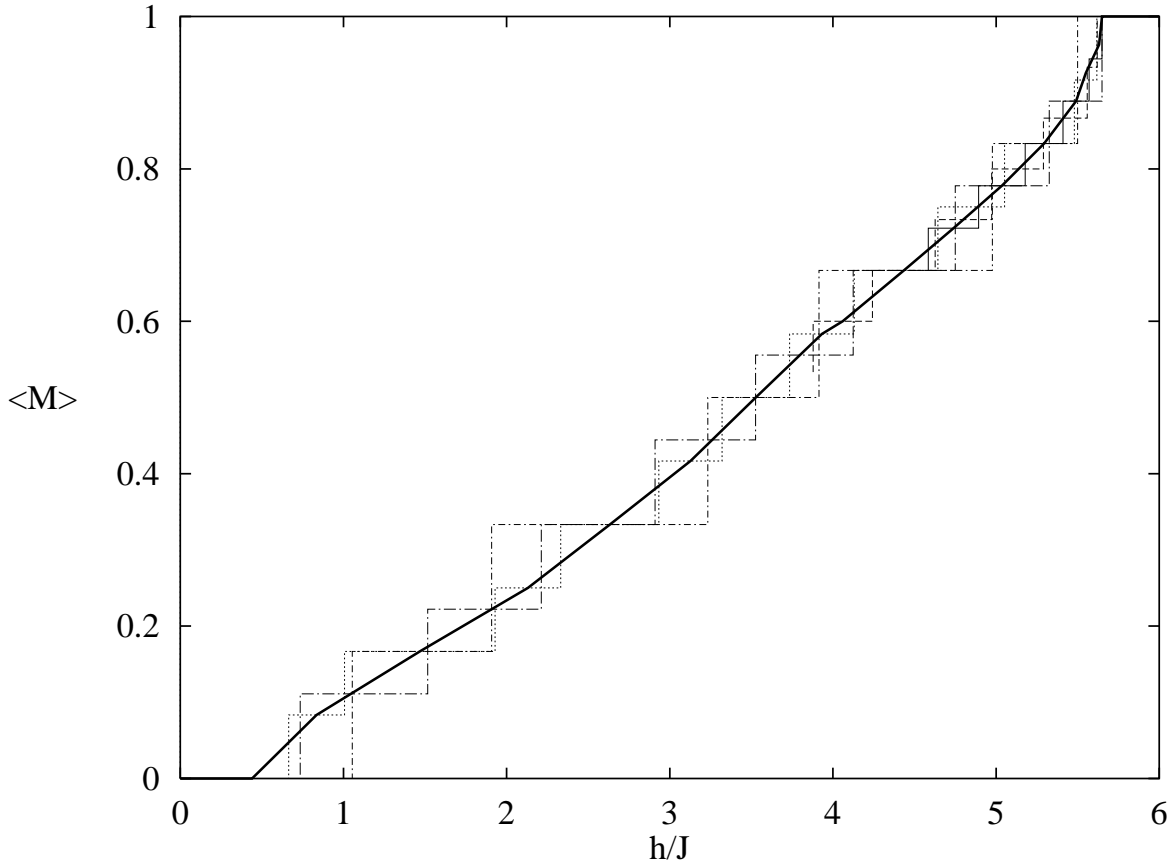,width=\columnwidth,angle=270}
\smallskip
\caption{
Same as Fig.\ \ref{fig4}, but for PBC of type B.
\label{fig5}
}
\end{figure}

\begin{figure}[ht]
\psfig{figure=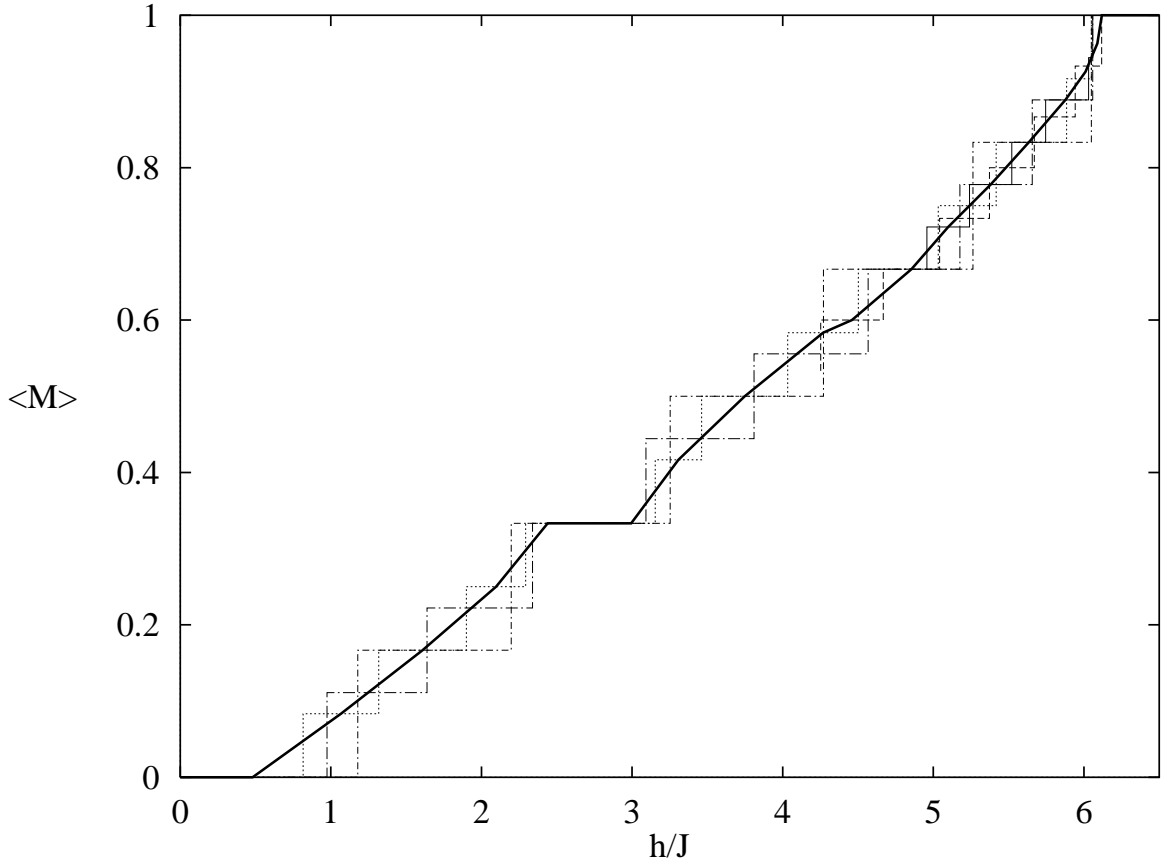,width=\columnwidth,angle=270}
\smallskip
\caption{
Same as Fig.\ \ref{fig4}, but for PBC of type C.
\label{fig6}
}
\end{figure}

Inspection of the $r=2$ version of (\ref{latHamN3pbcF}) shows that
at $J=0$, the $N=3$-leg zig-zag ladder with PBC of type B is equivalent to
the $N=2$-leg ladder with equal coupling constants.
So, the results for the two-leg case discussed earlier can be carried
over to this three-leg ladder at strong
coupling. For example, this $N=3$-leg ladder should exhibit incommensurate
groundstate momenta at $\Jp \gg J$. More important for our
purposes is that a small gap, but no other non-trivial plateaux
are expected at strong coupling. To see to which extent this
is generic, we show in Fig.\ \ref{fig5} numerical results obtained for
the smaller value of $\Jp = 3 J / 2$. Fig.\ \ref{fig6} shows the
analogous result for PBC of type C.

Since at $J=\Jp$, neither PBC of type B nor PBC of type C respect the
sublattice structure of the triangular lattice, there is no reason
to expect those $L$ which are a multiple of three to play any
particular r\^ole. We therefore simply consider even $L$.

If one ignores the bold extrapolated curves in Figs.\ \ref{fig5}
and \ref{fig6}, the finite-size magnetization curves look quite
similar at least if compared to the two boundary conditions
discussed before. Such a similarity of PBC of type B and type C
is also suggested by the fact that both of them can be mapped
to a single chain (\ref{latHamN3pbcF}). However, there are
definitely at least quantitative differences, as already
the value of the upper critical field shows:
At $\Delta=1$, $\Jp/J = 3/2$, one finds from (\ref{hucFpbc}) that
$h_{\rm uc} / J = 5 + {3 \over 8} \sqrt{3} \approx 5.6495$
for PBC of type B, while minimization of (\ref{E1sFpbc}) with
$r=4$ yields $h_{\rm uc} \approx 6.1193 J$ for PBC of type C.

The finite-size data for $L=4$, $6$ and $8$ indicates a small spin-gap
at zero field. For PBC of type B, application of a $1/L$-extrapolation
in the spirit of (\ref{1overLcorr}) leads
to a gap $\Gap \approx 0.24 J$ while the Shanks transform
yields $\Gap \approx 0.64 J$. Though both methods agree on the presence
of a gap, we are not able to determine it accurately -- by comparison
we conclude $\Gap/J = 0.44 \pm 0.20$ for PBC of type B. Similarly,
for PBC of type C, the Shanks transform yields $\Gap \approx 0.27 J$
while from (\ref{1overLcorr}) one finds $\Gap \approx 0.48 J$. Here, we
use the latter value though a large uncertainty should be kept in mind.

While one observes clear plateaux with $\langle M \rangle \ne 0$
in Figs.\ \ref{fig3} and \ref{fig4}, it is not immediately clear
from the finite-size data of Figs.\ \ref{fig5} and \ref{fig6} if
such non-trivial plateaux survive the thermodynamic limit.
This issue therefore requires a
more detailed discussion. First we take a closer look at
$\langle M \rangle = 1/3$. For PBC of type B, the finite-size plateau
with $L=4$, $6$ and $8$ is roughly (but not very well)
fitted by (\ref{ConfInf}).
More strongly, one finds a negative plateau width if one tries
to apply the extrapolation formula (\ref{1overLcorr}) to its
boundaries. This indicates that there is not even a tiny plateau
with $\langle M \rangle = 1/3$ in the thermodynamic limit $L \to \infty$.
For PBC of type C in contrast, the plateau width at $L=8$ is actually
larger than that at $L=6$. We have therefore drawn an $\langle M \rangle = 1/3$
plateau in Fig.\ \ref{fig6} whose boundaries have been determined using
(\ref{1overLcorr}).

At $\langle M \rangle = 2/3$ we have data
for a further system size ($L=10$). If we discard the $L=4$ data
and ignore some apparent non-monotonic finite-size effects we
actually obtain fairly good agreement with (\ref{ConfInf}) in
both cases. So, we do not find a plateau at $\langle M \rangle = 2/3$
in either of the two cases.

Extrapolated magnetization curves are shown by the bold lines in
Figs.\ \ref{fig5} and \ref{fig6}. They have some remaining wiggly
features which are related to the fact that non-monotonic finite-size
effects may still be important at the system sizes used for the
extrapolation.

The absence of a plateau e.g.\ at $\langle M \rangle = 1/3$ can be
attributed to enhanced translational symmetry: The $N=3$ zig-zag
ladder with PBC of type B can be mapped to a single
Heisenberg chain (\ref{latHamN3pbcF}). A plateau
at $\langle M \rangle = 1/3$ would then require breaking of the
enhanced translational symmetry by $l=3$ sites and a plateau at
$\langle M \rangle = 2/3$ would require the even larger period
of $l=6$ sites (compare (\ref{condM}) where $V=l$ due to the
enhanced symmetry). However, to the best of our knowledge
spontaneous symmetry breaking with periods of more than two sites
has not been observed so far (at least for ordinary ladders it
is unlikely to occur \cite{CHP2}).

It should be mentioned though that the same argument can be applied
to PBC of type C, where a plateau with $\langle M \rangle = 1/3$ seems
likely. Furthermore, all versions of PBC are very similar from
the weak-coupling point of view (compare Section III). In particular, one
would expect a plateau with $\langle M \rangle = 1/3$ for all versions of
PBC if it appears for one of them in the weak-coupling region.
Which plateaux are actually absent and which ones present therefore
needs further investigation.

\section{Discussion and Conclusion}

In this paper we have analyzed zig-zag coupled chains using a
range of field-theoretical and numerical techniques. First, we have
discussed the case of zero field and weak interchain coupling where
in a field-theoretical formulation the most relevant interaction for
the ordinary ladders is replaced with a chirally asymmetric one \cite{NGE}.
In the $SU(2)$ symmetric case this chirally asymmetric term is marginal
and then current-current interaction terms (usually not considered in
the presence of the relevant terms that arise in the normal ladders)
have to be included in the RG analysis, leading us to our one-loop RG
equations (\ref{RGeq2}). In this approximation, one sees that the effect
of the chirally asymmetric term is to push the system further into the
massive phase for the AF case, and to open a gap for small $\Jp < 0$.
The same is true for $N\ge 3$, when all the chains are (weakly)
antiferromagnetically coupled in an equal manner. For sufficiently
small $\Delta <1$, the zero magnetization groundstate is a massless
$c=1$ theory, in contrast to the $N=2$ ladder with normal couplings.

We have then shown that this more relevant interaction is restored by
a magnetic field, dimerization or doping with charge
carriers. The latter variants of zig-zag ladders should therefore
be similar to the usual spin ladders. We have also analyzed the
appearance of plateaux in the magnetization curves of $N\ge 3$ zig-zag
ladders. We have found that the situation is
similar to that encountered in the normal ladders (in the weak coupling
limit) \cite{CHP2}, except for a trivial rescaling of the couplings which
lead to slightly different quantitative predictions for the opening points.

These considerations may also be relevant to the description of
experiments. Firstly, we have seen that under very general conditions
the normal perpendicular couplings are those that really matter.
Secondly, we have argued that dimerization along the legs (see
Fig.\ \ref{figdim2}) might be an important feature in understanding
the magnetization experiments \cite{STKT} on NH$_4$CuCl$_3$
(see also \cite{CG2} for a related observation in the context
of the usual spin ladders with dimerization along the chains).
In the present context, this
is suggested by the limit of strong dimerization $\delta \to 1$
which we have discussed in the appendix. Dimerization also seems
to be a small enough modification to be plausible,
but this will have to be confirmed by a determination
of the structure of NH$_4$CuCl$_3$ at low temperatures.

The field-theoretical approach is complemented in a second part
by a numerical analysis. First, we have analyzed a two-leg
zig-zag ladder without dimerization. For positive $\Jp$ we have
confirmed known facts such as the appearance of a spin-gap but no
non-trivial plateaux in the magnetization curve. The magnetized
groundstates exhibit two different types of behaviour:
At small $\Jp$ incommensurate groundstate momenta participate
in the magnetization process, while for small $J$ all momenta
are commensurate. For the case of ferromagnetic coupling $\Jp < 0$
we have found interesting new behaviour: At small $\abs{\Jp}$
spins flip in pairs and for this reason all magnetized groundstates
are commensurate. At larger $\abs{\Jp}$ more than two spins can
flip simultaneously and at the same time also incommensurate groundstate
momenta become relevant to the magnetization process.
Finally, for $\Jp < -4 J$, the behaviour is
completely ferromagnetic. The `phases' with incommensurate
groundstate momenta are reminiscent of similar phenomena observed
in a spin-one chain \cite{FaLi,GJS}. It has been conjectured \cite{FaLi,FraRoe}
that also the two-leg zig-zag ladder gives rise to a two-component
Luttinger liquid, but at least in the weak-coupling region we did not
find evidence for more than one massless degree of freedom.

We have then investigated the transition to saturation in more
detail and confirmed the univeral DN-PT behaviour
(\ref{DNPT}). The only exception is $\Jp = 4J$ where a
quartic behaviour (\ref{nDNPT}) was already observed in \cite{SGMK}.
This modified exponent has a simple explanation in terms of
an exceptional behaviour of the single-spinwave dispersion
which appears not to have been noted before. In the region
of not too strong ferromagnetic $\Jp < 0$ spins flip in pairs
and this relation to the single-spinwave dispersion is lost.
It is therefore non-trivial that our numerical data for the asymptotics
of the magnetization curve is still consistent with the DN-PT universal
behaviour.

Finally, we have computed magnetization curves for four
different variants of a three-leg ladder. We observed a remarkable
dependence on the geometry of the interchain coupling
(see Figs.\ \ref{fig3}--\ref{fig6}).
Clear plateaux can be observed in several cases while indications
against the presence of such plateaux were obtained in other situations
\footnote{It should be noted that due to non-monotonic finite-size
effects it is difficult to reliably exclude plateaux not only in the
cases discussed in Section VI but also otherwise. Nevertheless,
the evidence for or against a plateau e.g. at $\langle M \rangle = 1/3$
for $N=3$ depends strongly on the boundary conditions.}.
We expect that the same types of commensurate and incommensurate
phases as observed for the two-leg ladder are also present in
zig-zag ladders with three or more legs.
This is indicated e.g.\ by the analysis of single-spinwave excitations
for three-leg ladders, but would need verification. Further new
phenomena might appear for more than two coupled chains.

All the observed plateaux can be interpreted in terms of
the quantization condition (\ref{condM}). The essence is
enhanced translational symmetry in certain cases
which induced by frustration is then spontaneously broken
(in most cases to periods $l \le 2$). Specifically, for
both the ordinary and the zig-zag two-leg ladder at $\delta = 0$
one then has a translationally invariant unit cell
containing $V=2$ spins. However, the interpretation
is different: For zig-zag coupling, translational symmetry is
first enhanced by a factor of two and then spontaneosly broken by a period
$l=2$. This also explains why one needs a dimerized interchain
coupling ($0 < \delta < 1$) to have a plateau with $\langle M \rangle = 1/2$
in the two-leg zig-zag ladder \cite{Totsuka2,TNK,FGKMW} which is
permitted by (\ref{condM}) with $N=2$, $l=2$ only in the presence
of dimerization. Dimerization along the legs breaks translational symmetry
further to $V = 4 l$. Frustration-induced spontaneous symmetry
breaking to a period $l=2$ then permits the aforementioned
appearance of further plateaux with $\langle M \rangle = 1/4$ and
$3/4$.

To interpret our results for the three-leg zig-zag ladders in terms
of (\ref{condM}), one should substitute $V=3 l$ for OBC and PBC of
type A and $V = l$ for PBC of type B and C. With one exception,
all the plateaux sketched in Figs. \ref{fig3}--\ref{fig6} can then
be naturally interpreted with periods $l \le 2$. Only
the plateau with $\langle M \rangle = 1/3$ for PBC of type C
in Fig,\ \ref{fig6} requires a period $l=3$ which (if the
presence of this plateau is confirmed) would be the highest
observed period which we are aware of.

In fact, the numerical support for a plateau with $\langle M \rangle = 1/3$
for $N=3$ and PBC of type C is rather weak and would deserve
further attention. Conversely,
the absence of a spin-gap in Fig.\ \ref{fig3} is not on safe grounds,
and the Abelian bosonization analysis would actually predict at least
a small spin-gap in the weak-coupling region. Another issue for $N=2$
which requires further attention is that there is no numerical evidence yet
for the spin-gap predicted by field theory for weak ferromagnetic $\Jp < 0$.

So far, we have just observed numerically for $N=3$ that modifications
in the boundary conditions have a drastic effect on the magnetization
process. This is in contrast to ordinary spin ladders \cite{CHP2}
and not apparent in the field theoretical treatment. It may
be necessary to include higher loop corrections or to perform a
non-perturbative analysis of the RG equations in order
to understand how the novel interaction term arising from the
zig-zag coupling gives for example rise to the differences in the
different versions of PBC.

In summary, on the one hand we believe that we have exhibited
interesting properties of zig-zag spin ladders. On the other hand,
there is a number of points which deserve further attention. We hope that
with this combination, the present paper will stimulate further
research on zig-zag spin ladders.

\acknowledgments
We are indebted to A.A.\ Nersesyan for many useful discussions and
a critical reading of the manuscript. We would also like to thank
K.\ Le Hur, P.\ Lecheminant and M.E.\ Zhitomirsky for useful discussions
and comments.
D.C.C.\ would like to acknowledge financial support from CONICET,
Fundaci\'on Antorchas, Deutsche Ausgleichbank, ANPCyT
(under grant No.\ 03-00000-02249), the DAAD (under the Visiting
Professors Programme) and thank the Physikalisches Institut der Universit\"at
Bonn for hospitality. The more involved numerical computations have been
carried out on computers of the Max-Planck-Institut f\"ur Mathematik,
Bonn-Beuel.

\appendix
\section{Strong-coupling effective Hamiltonian for
a dimerized two-leg ladder}

In this appendix we
consider the Hamiltonian for a two-leg zig-zag ladder with
dimerized chains and coupling between the chains:
\bea
H &=& J \sum_{i=1}^2 \sum_{x=1}^L
      \left(1 + (-1)^x \delta\right) \Sv_{i,x} \cdot \Sv_{i,x+1} \nn \\
&& + \Jp \sum_{x=1}^L \left\{ (1+\deltap) \Sv_{1,x} \cdot \Sv_{2,x}
      + (1-\deltap) \Sv_{1,x} \cdot \Sv_{2,x+1} \right\} \label{Hddim} \\
&& - h \sum_{i=1}^2 \sum_{x=1}^L \Sz_{i,x} \, . \nn
\eea
If we assume that $\Jp (1 \pm \deltap), J (1-\delta) \ll J (1+\delta)$,
we can describe the transition from $\langle M \rangle = 0$
to $\langle M \rangle = 1$ by an effective Hamiltonian following
\cite{CJYFHBLHP,Totsuka2,Mila,TLPRS,Totsuka3,FZ}. On
each `rung' (a bond coupled with coefficient $J (1+\delta)$)
we retain only two states: The singlet and the fully polarized
state. With an appropriate choice of basis, the first-order effective
Hamiltonian can then be written as (note that we include
the external magnetic field $h$ in zeroth order):
\bea
H_{\rm eff.} &=&
J \left(1 - \delta\right) \sum_{x=1}^L \left\{
     {1 \over 4} \left(S_x^{+} S_{x+2}^{-} + S_x^{-} S_{x+2}^{+} \right)
   + {1 \over 4} \Sz_x \Sz_{x+2}
   \right\} \nn \\
&& + \Jp \left(1 - \deltap\right) \sum_{x=1}^{L/2} \left\{
     {1 \over 4} \left(S_{2x}^{+} S_{2x+1}^{-} + S_{2x}^{-} S_{2x+1}^{+} \right)
   + {1 \over 4} \Sz_{2x} \Sz_{2x+1}
   \right\} \nn \\
&& + \Jp \left(1 + 3 \deltap\right) \sum_{x=1}^{L/2}
     {1 \over 4} \left(S_{2x+1}^{+} S_{2x+2}^{-}
                     + S_{2x+1}^{-} S_{2x+2}^{+} \right)
   + \Jp \left(3 + \deltap\right) \sum_{x=1}^{L/2}
     {1 \over 4} \Sz_{2x+1} \Sz_{2x+2} \nn \\
&& + \left({1 \over 4} J (1-\delta) + {\Jp \over 2}\right)
    \left({L \over 4} + \sum_{x=1}^L \Sz_x \right) \, .
\label{HddEff}
\eea
The first line comes from one weak coupling $J \left(1 - \delta\right)$
along the two original chains, the second one from one coupling
$\Jp \left(1 - \deltap\right)$ between the two chains and the
third line arises from one coupling $\Jp \left(1 - \deltap\right)$
plus two couplings $\Jp \left(1 + \deltap\right)$ between the
chains. The fourth line is just a first-order correction to the
external magnetic field (apart from a trivial additive constant).

In eq.\ (\ref{HddEff}) one recognizes again a two-leg zig-zag ladder with
dimerized coupling between the chains. The coupling
along the two chains is $J (1-\delta)$ with an effective $XXZ$ anisotropy
$\Delta_{\rm eff.} = {1/2}$. The coupling between the
two chains is not only dimerized but also has an alternating $XXZ$ anisotropy:
$\Jpeff = \Jp (1-\deltap)$, $\Delta'_{\rm eff.} = {1/2}$
on even sites, $\Jpeff = \Jp (1 + 3 \deltap)$,
$\Delta'_{\rm eff.} = (3 + \deltap)/(2 (1 + 3 \deltap))$ on the odd
sites. For these values of parameters, the two-leg zig-zag ladder
in a magnetic field has unfortunately not yet been studied in detail.
However, this mapping is still suggestive since we know that a two-leg
zig-zag ladder with dimerized coupling between the chains can exhibit
plateaux at $\langle M_{\rm eff.} \rangle = 0$ and
$\langle M_{\rm eff.} \rangle = \pm {1 /2}$ \cite{Totsuka2,TNK,FGKMW}.
This suggests the possibility of the two-leg
ladder with dimerized coupling along the chains (\ref{Hddim}) having not
only plateaux at $\langle M \rangle = 0$ and $\langle M \rangle = {1 / 2}$
(corresponding to $\langle M_{\rm eff.} \rangle = 0$)
but also at $\langle M \rangle = {1 / 4}$ ($\langle M_{\rm eff.} \rangle
= -{1 / 2}$) and $\langle M \rangle = {3 / 4}$
($\langle M_{\rm eff.} \rangle = {1 / 2}$). This observation may be
relevant to the magnetization experiments on NH$_4$CuCl$_3$ \cite{STKT},
since these are precisely the observed magnetization plateaux.
However, in the strong dimerization limit considered in this appendix,
one would certainly have a pronounced $\langle M \rangle = 0$ plateau
which is not observed in NH$_4$CuCl$_3$. So, even if dimerization
along the chains should be important in this compound, it
cannot really lie in the region where the above mapping is applicable.


\begin{references}
\bibitem{review} E.\ Dagotto, T.M.\ Rice, Science {\bf 271}, 618 (1996);
              T.M.\ Rice, Z.\ Phys.\ {\bf B103}, 165 (1997).
\bibitem{Ptoday} B.G.\ Levi, Physics Today, October 1996, p.\ 17.
\bibitem{CJYFHBLHP} G.\ Chaboussant, M.-H.\ Julien, Y.\ Fagot-Revurat,
              M.\ Hanson, C.\ Berthier, L.P.\ L\'evy, M.\ Horvati\'c,
              O.\ Piovesana, Eur.\ Phys.\ J.\ {\bf B6}, 167 (1998).
\bibitem{Hida} K.\ Hida, J. Phys. Soc. Jpn. {\bf 63}, 2359 (1994);
              K.\ Okamoto, Solid State Comm.\ {\bf 98}, 245 (1995).
\bibitem{Tone} T.\ Tonegawa, T. Nakao, M. Kaburagi,
              J.\ Phys.\ Soc.\ Jpn.\ {\bf 65}, 3317 (1996).
\bibitem{AOY} M.\ Oshikawa, M.\ Yamanaka, I.\ Affleck,
              Phys.\ Rev.\ Lett.\ {\bf 78}, 1984 (1997).
\bibitem{CCLMMP} G.\ Chaboussant, P.A.\ Crowell, L.P.\ L\'evy,
             O.\ Piovesana, A.\ Madouri, D.\ Mailly, Phys.\ Rev.\ {\bf B55},
             3046 (1997).
\bibitem{HLP} C.A.\ Hayward, D.\ Poilblanc, L.P.\ L\'evy,
              Phys.\ Rev.\ {\bf B54}, R12649 (1996).
\bibitem{Totsuka} K.\ Totsuka, Phys.\ Lett.\ {\bf A228}, 103 (1997).
\bibitem{CHP} D.C.\ Cabra, A.\ Honecker, P.\ Pujol,
              Phys.\ Rev.\ Lett.\ {\bf 79}, 5126 (1997).
\bibitem{SaTaS3o2} T.\ Sakai, M.\ Takahashi, Phys.\ Rev.\ {\bf B57}, R3201
             (1998).
\bibitem{Totsuka2} K.\ Totsuka, Phys.\ Rev.\ {\bf B57}, 3454 (1998).
\bibitem{CHP2} D.C.\ Cabra, A.\ Honecker, P.\ Pujol, Phys.\ Rev.\ {\bf B58},
              6241 (1998).
\bibitem{CG} D.C.\ Cabra, M.D.\ Gryn\-berg, Phys.\ Rev.\ {\bf B59}, 119 (1999).
\bibitem{H} A.\ Honecker, Phys.\ Rev.\ {\bf B59}. 6790 (1999).
\bibitem{Mila} F.\ Mila, Eur.\ Phys.\ J.\ {\bf B6}, 201 (1998).
\bibitem{TLPRS} K.\ Tandon, S.\ Lal, S.K.\ Pati, S.\ Ramasesha, D.\ Sen,
              Phys.\ Rev.\ {\bf B59}, 396 (1999).
\bibitem{Totsuka3} K.\ Totsuka, Eur.\ Phys.\ J.\ {\bf B5}, 705 (1998).
\bibitem{FZ} A.\ Furusaki, S.C.\ Zhang, 
              Phys.\ Rev.\ {\bf B60}, 1175 (1999).
\bibitem{CG2} D.C.\ Cabra, M.D.\ Gryn\-berg, Phys.\ Rev.\ Lett.\ {\bf 82},
              1768 (1999).
\bibitem{CPKSR} R.\ Chitra, S.K.\ Pati, H.R.\ Krishnamurthy, D.\ Sen,
              S.\ Ramasesha, Phys.\ Rev.\ {\bf B52}, 6581 (1995).
\bibitem{WA} S.R.\ White, I.\ Affleck, Phys.\ Rev.\ {\bf B54}, 9862 (1996).
\bibitem{AS} D.\ Allen, D.\ S\'en\'echal, Phys.\ Rev.\ {\bf B55}, 299 (1997).
\bibitem{NGE} A.A.\ Nersesyan, A.O.\ Gogolin, F.H.L.\ E{\ss}ler,
              Phys.\ Rev.\ Lett.\ {\bf 81}, 910 (1998).
\bibitem{Sor} E.\ S{\o}rensen, I.\ Affleck, D.\ Augier, D.\ Poilblanc,
              Phys.\ Rev.\ {\bf B58}, R14701 (1998).
\bibitem{CTC} R.\ Coldea, D.A.\ Tennant, R.A.\ Cowley, D,F.\ McMorrow,
              B.\ Dorner, Z.\ Tylczynski, J.\ Phys.\ Cond.\ Mat.\ {\bf 8},
              7473 (1996); Phys.\ Rev.\ Lett.\ {\bf 79}, 151 (1997).
\bibitem{STKTKTMG} W.\ Shiramura, K.\ Takatsu, H.\ Tanaka, K.\ Kamishima,
              M.\ Takahashi, H.\ Mitamura, T.\ Goto, J.\ Phys.\ Soc.\ Jpn.\
              {\bf 66}, 1900 (1997).
\bibitem{STKT} W.\ Shiramura {\it et al}, J.\ Phys.\ Soc.\ Jpn.\ {\bf 67},
              1548 (1998).
\bibitem{Kolezhuk} A.K.\ Kolezhuk, Phys.\ Rev.\ {\bf B59}, 4181 (1999).
\bibitem{AlAu} E.\ Altman, A.\ Auerbach, Phys.\ Rev.\ Lett.\ {\bf 81}, 4484
              (1998).
\bibitem{YHMM} S.\ Yunoki, J.\ Hu, A.L.\ Malvezzi, A.\ Moreo, N.\ Furukawa,
              E.\ Dagotto, Phys.\ Rev.\ Lett.\ {\bf 80}, 845 (1998).
\bibitem{MaGo} C.K.\ Majumdar, D.K.\ Ghosh, J.\ Math.\ Phys.\ {\bf 10},
              1388 (1969);
              J.\ Math.\ Phys.\ {\bf 10}, 1399 (1969).
\bibitem{Maju} C.K.\ Majumdar, J.\ Phys.\ C: Solid State Phys.\ {\bf 3}, 911
              (1970).
\bibitem{ShSu} B.S.\ Shastry, B.\ Sutherland, Phys.\ Rev.\ Lett.\ {\bf 47},
              964 (1981).
\bibitem{Wi} E.\ Witten, Commun.\ Math.\ Phys.\ {\bf 92}, 455 (1984).
\bibitem{A2} I.\ Affleck,
              in {\it Fields, Strings and Critical Phenomena, Les Houches,
              Session XLIX}, edited by E.\ Br\'ezin and J.\ Zinn-Justin
              (North-Holland, Amsterdam, 1988).
\bibitem{HF} T.\ Hikihara, A.\ Furusaki, Phys.\ Rev.\ {\bf B58}, R583 (1998).
\bibitem{LZ} S.\ Lukyanov, A.\ Zamolodchikov, Nucl.\ Phys.\ {\bf B493},
              571 (1997).
\bibitem{TNK} T.\ Tonegawa, T.\ Nishida, M.\ Kaburagi,
              Physica {\bf B246-247}, 368 (1998).
\bibitem{FGKMW} A.\ Fledderjohann, C.\ Gerhardt, M.\ Karbach, K.-H.\ M\"utter,
              R.\ Wie{\ss}ner, Phys.\ Rev.\ {\bf B59}, 991 (1999).
\bibitem{ST} D.G.\ Shelton, A.M.\ Tsvelik, Phys.\ Rev.\
              {\bf B53}, 14036 (1996).
\bibitem{FK} H.\ Frahm, V.E.\ Korepin, Phys.\ Rev.\ {\bf B42}, 10533 (1990).
\bibitem{ToHa1} T.\ Tonegawa, I.\ Harada, J.\ Phys.\ Soc.\ Jpn.\ {\bf 56},
              2153 (1987).
\bibitem{ToHa2} T.\ Tonegawa, I.\ Harada, Physica {\bf B155}, 379 (1989).
\bibitem{ToHa3} T.\ Tonegawa, I.\ Harada, J.\ Phys.\ Soc.\ Jpn.\ {\bf 58},
              2902 (1989).
\bibitem{SGMK} M.\ Schmidt, C.\ Gerhardt, K.-H.\ M\"utter, M.\ Karbach,
              J.\ Phys.: Condensed Matter {\bf 8}, 553 (1996).
\bibitem{GFAMAK} C.\ Gerhardt, A.\ Fledderjohann, E.\ Aysal, K.-H.\ M\"utter,
              J.F.\ Audet, H.\ Kr\"oger, J.\ Phys.: Condensed Matter {\bf 9},
              3435 (1997).
\bibitem{GMK} C.\ Gerhardt, K.-H.\ M\"utter, H.\ Kr\"oger, Phys.\ Rev.\ {\bf
              B57}, 11504 (1998).
\bibitem{UsSu} M.\ Usami, S.\ Suga, Phys.\ Lett.\ {\bf A240}, 85 (1998).
\bibitem{FaLi} G.\ F\'ath, P.B.\ Littlewood, Phys.\ Rev.\ {\bf B58},
              R14709 (1998).
\bibitem{GJS} O.\ Golinelli, Th.\ Jolicoeur, E.S.\ S{\o}rensen,
              Eur.\ Phys.\ J.\ {\bf B11}, 199 (1999).
\bibitem{BGFPXZ} R.J.\ Bursill, G.A.\ Gehring, D.J.J.\ Farnell, J.B.\ Parkinson,
              T.\ Xiang, C.\ Zeng, J.\ Phys.: Condensed Matter {\bf 7}, 8605
              (1995).
\bibitem{FraRoe} H.\ Frahm, C.\ R\"odenbeck,
               Eur.\ Phys.\ J.\ {\bf B10}, 409 (1999).
\bibitem{HKNN} T.\ Hamada, J.\ Kane, S.\ Nakagawa, Y.\ Natsume, J.\ Phys.\
              Soc.\ Jpn.\ {\bf 57}, 1891 (1988).
\bibitem{BoFi} J.C.\ Bonner, M.E.\ Fisher, Phys Rev.\ {\bf 135}, A640 (1964).
\bibitem{Cardy} J.L.\ Cardy, J.\ Phys.\ A: Math.\ Gen.\ {\bf 17}, L385 (1984).
\bibitem{HoPa} R.P.\ Hodgson, J.B.\ Parkinson, J.\ Phys.\ C: Solid
              State Phys.\ {\bf 18}, 6385 (1985).
\bibitem{DzNe} G.I.\ Dzhaparidze, A.A.\ Nersesyan,
              JETP Lett.\ {\bf 27}, 334 (1978).
\bibitem{PoTa} V.L.\ Pokrovsky, A.L.\ Talapov,
              Phys.\ Rev.\ Lett.\ {\bf 42}, 65 (1979).
\bibitem{MiNi} H.\ Nishimori, S.\ Miyashita, J.\ Phys.\ Soc.\ Jpn.\ {\bf 55},                 4448 (1986).
\bibitem{twodim} A.\ Honecker, J.\ Phys.: Condensed Matter {\bf 11},
              4697 (1999).
\bibitem{KaTa} K.\ Kawano, M.\ Takahashi,
              J.\ Phys.\ Soc.\ Jpn.\ {\bf 66}, 4001 (1997).
\bibitem{SaTa} T.\ Sakai, M.\ Takahashi,
              Phys.\ Rev.\ {\bf B43}, 13383 (1991);
              J.\ Phys.\ Soc.\ Jpn.\ {\bf 60}, 3615 (1991).
\bibitem{SMWSD} G.\ Sierra, M.A.\ Mart\'{\i}n-Delgado, S.R.\ White,
             D.J.\ Scalapino, J.\ Dukelsky, Phys.\ Rev.\ {\bf B59}, 7973 (1999).
\bibitem{Kuramoto} T.\ Kuramoto, J.\ Phys.\ Soc.\ Jpn.\ {\bf 67}, 1762
             (1998).
\bibitem{MSBMY} K.\ Maisinger, U.\ Schollw\"ock, S.\ Brehmer, H.-J.\ Mikeska,
             S.\ Yamamoto, Phys.\ Rev.\ {\bf B58}, R5908 (1998).
\bibitem{HeSchue} M.\ Henkel, G.M.\ Sch\"utz, J.\ Phys.\ A: Math.\ Gen.\
              {\bf 21}, 2617 (1988).
\bibitem{ALN} P.\ Azaria, P.\ Lecheminant, A.A.\ Nersesyan,
             Phys.\ Rev.\ {\bf B58}, R8881 (1998).

\end{references}
\end{document}